\documentclass[twoside,11pt]{article}
\usepackage{jmlr2e}

\usepackage{times,url}
\usepackage{amsmath}
\usepackage{amsfonts}
\usepackage{graphicx}
\usepackage{subfigure}

\ShortHeadings{Cauchy Random Projections}{Li, Hastie, and Church}
\firstpageno{1}

\begin{document}
\title{Nonlinear Estimators and Tail Bounds for Dimension Reduction in
  $l_1$ Using Cauchy Random Projections}

\author{\name Ping Li \email pingli@stat.stanford.edu \\
       \addr Department of Statistics\\
       Stanford University\\
       Stanford, CA 94305, USA
       \AND
       \name Trevor J.\ Hastie \email hastie@stanford.edu \\
       \addr Department of Statistics\\
       Stanford University\\
       Stanford, CA 94305, USA
       \AND
       \name Kenneth W.\ Church \email church@microsoft.com \\
       \addr Microsoft Research\\
       Microsoft Corporation\\
       Redmond, WA 98052, USA
}
\editor{}

\maketitle
\vspace{-0.5in}
\begin{abstract}

For \footnote{Revised \today. The original version, titled {\em
    Practical Procedures for Dimension Reduction in $l_1$}, is 
  available as a technical report in Stanford Statistics achive
  (report No. 2006-04, June, 2006). } 
 dimension reduction in $l_1$, the method of {\em Cauchy random projections} multiplies
the original data matrix $\mathbf{A} 
\in\mathbb{R}^{n\times D}$ with a random matrix $\mathbf{R} \in
\mathbb{R}^{D\times k}$ ($k\ll\min(n,D)$) whose entries are i.i.d. samples
of the standard Cauchy $C(0,1)$. Because of the impossibility results,  one can not
hope to recover the pairwise $l_1$ distances in $\mathbf{A}$
from $\mathbf{B} = \mathbf{AR} \in \mathbb{R}^{n\times k}$, 
using linear estimators without incurring large
errors. However, nonlinear estimators are still useful for certain
applications in data stream computation, information
retrieval, learning, and data mining.  

We propose three types of nonlinear estimators: the bias-corrected
sample median estimator, the bias-corrected geometric mean estimator,
and the bias-corrected maximum likelihood estimator. The sample median
estimator and the geometric mean estimator are asymptotically (as
$k\rightarrow \infty$) equivalent but the latter is more accurate at
small $k$.  We derive explicit tail bounds
for the geometric mean estimator and establish an analog of the
Johnson-Lindenstrauss  (JL) lemma for dimension reduction in $l_1$,
which is weaker than the classical JL lemma for dimension reduction in
$l_2$. 

Asymptotically, both the sample median estimator and the
geometric mean estimators are about $80\%$ efficient compared to the
maximum likelihood estimator (MLE). We analyze the moments of the MLE
and propose approximating the distribution of the MLE by an 
inverse Gaussian. 

\end{abstract}


\section{Introduction}

This paper focuses on dimension reduction in $l_1$,  in particular, on
the
method based on {\em Cauchy random projections}
\citep{Proc:Indyk_FOCS00}, which is special case of {\em linear
  random projections}. 

The idea of {\em linear random projections} is to multiply the original data
matrix $\mathbf{A} \in \mathbb{R}^{n\times D}$ with a random projection matrix
$\mathbf{R} \in \mathbb{R}^{D\times k}$, resulting in a projected
matrix $\mathbf{B} = \mathbf{AR} \in \mathbb{R}^{n\times
  k}$. If $k \ll \min(n,D)$, then it should be much more efficient to
compute certain summary statistics (e.g., pairwise distances) from 
$\mathbf{B}$ as opposed to $\mathbf{A}$. Moreover, $\mathbf{B}$ may be small 
enough to reside in physical memory while $\mathbf{A}$ is often
too large to fit in the main memory.  

The choice of the random projection matrix $\mathbf{R}$ depends on which norm we
would like to work with. 
\cite{Proc:Indyk_FOCS00} proposed constructing $\mathbf{R}$ from 
i.i.d. samples of $p$-stable distributions, for dimension reduction in
$l_p$ ($0< p\leq 2$). In the stable distribution family \citep{Book:Zolotarev_86}, normal is
2-stable and Cauchy is 1-stable. Thus, we will call random projections
for $l_2$ and $l_1$,  {\em normal random projections} and {\em Cauchy
  random projections}, respectively. 

In {\em normal random projections} \citep{Book:Vempala}, we can estimate the original
pairwise $l_2$ distances of $\mathbf{A}$ directly using the
corresponding $l_2$ distances of $\mathbf{B}$ (up to a normalizing
constant). Furthermore, the Johnson-Lindenstrauss  (JL)
lemma \citep{Article:JL84} provides the performance guarantee.
 We will review {\em normal random projections} in more detail in
 Section \ref{sec_intr_rp}.

For {\em Cauchy random projections}, we should not use the $l_1$ distance
in $\mathbf{B}$ to approximate the original $l_1$ distance in
$\mathbf{A}$, as the Cauchy distribution does not even have  a finite first
moment. The impossibility results
\citep{Proc:Brinkman_FOCS03,Article:Lee_Naor_04,Article:Brinkman_JACM05}
have proved that one can not hope to recover the $l_1$ distance using
linear projections and linear estimators (e.g., sample mean), without
incurring large errors.  Fortunately, the impossibility results do not
 rule out nonlinear estimators, which may be still useful in
certain applications in data stream computation, information
retrieval, learning, and data mining. 

\cite{Proc:Indyk_FOCS00} proposed using the sample median (instead of
the sample mean) in {\em Cauchy random projections} and described its
application in data stream computation. In this study, we provide
three types of nonlinear estimators:  the bias-corrected
sample median estimator, the bias-corrected geometric mean estimator,
and the bias-corrected maximum likelihood estimator. The sample median
estimator and the geometric mean estimator are asymptotically
equivalent (i.e., both are about $80\%$ efficient as the maximum
likelihood estimator), but the latter is more accurate at small sample size $k$. 
Furthermore, we
derive explicit tail bounds for the bias-corrected geometric mean estimator and
establish an analog of the  JL Lemma for dimension reduction in
$l_1$. 

This analog of the JL Lemma for $l_1$ is weaker than the classical
JL Lemma for $l_2$, as the geometric mean estimator is a non-convex
norm and hence is not a metric.  Many efficient algorithms, such as
some sub-linear time (using super-linear memory) nearest neighbor algorithms \citep{Book:NN_05}, rely
on the metric properties (e.g., the triangle inequality). Nevertheless, nonlinear estimators may be
still useful in important scenarios. 
\begin{itemize}
\item {\em Estimating $l_1$ distances online} \\
The original data matrix $\mathbf{A} \in \mathbb{R}^{n\times D}$
requires $O(nD)$ storage space; and hence it is
often too large for physical memory. The storage cost of all
pairwise distances is $O(n^2)$, which may be also too large for the 
memory. For example, in information retrieval, $n$ could be
the total number of  word types or documents at Web scale. To avoid page fault,
it may be more efficient to estimate the distances on the fly from
the  projected data matrix $\mathbf{B}$ in the memory.  
\item {\em Computing all pairwise $l_1$ distances} \\
In distance-based clustering and  classification applications, we need
to compute all pairwise distances in $\mathbf{A}$, at the cost of
time $O(n^2D)$. Using {\em Cauchy random projections}, the cost can be reduced
to $O(nDk + n^2k)$. Because $k \ll \min(n,D)$, the savings
could be enormous. 
\item {\em Linear scan nearest neighbor searching}\\
We can always search for the nearest neighbors by linear scans. When
working with the projected data matrix $\mathbf{B}$ (which is in the  memory), the cost of
searching for the nearest neighbor for one data point is time $O(nk)$,
which may be still significantly faster than the sub-linear algorithms
working with the original data matrix $\mathbf{A}$ (which is often on the
disk). 
\end{itemize}

We briefly comment on {\em coordinate
  sampling}, another strategy for dimension reduction.  Given a data matrix $\mathbf{A}
\in \mathbb{R}^{n\times D}$, one can randomly sample $k$ columns from $\mathbf{A}$ and
estimate the summary statistics (including $l_1$ and $l_2$
distances). Despite its simplicity, there are two
major disadvantages in
coordinate sampling. First, there is no performance guarantee. For 
heavy-tailed data, we may have to choose $k$ very large in order to
achieve sufficient accuracy. Second, large datasets are often highly sparse,
for example,  text data \citep{Article:Dhillon_ML01} and market-basket
data \citep{Proc:Aggarwal_Wolf_Sigmod99,Proc:Strehl_HiPC00}.  \cite{Report:Li_Church_Sketch} and  \cite{Report:Li_Church_Hastie_crs}
provide an alternative coordinate sampling strategy, called
{\em Conditional Random Sampling (CRS)}, suitable for sparse
data. For non-sparse data, however, methods based on {\em linear 
  random projections} are superior. 

The rest of the paper is organized as follows. Section \ref{sec_intr_rp}
reviews {\em linear random projections}. Section \ref{sec_results}
summarizes the main results for three types of nonlinear
estimators. Section \ref{sec_median} presents the sample median
estimators. Section \ref{sec_gm} concerns the geometric mean
estimators. Section \ref{sec_mle} is devoted to the maximum likelihood
estimators. Section \ref{sec_conclusion}
concludes the paper.  

\section{Introduction to Linear Random Projections}\label{sec_intr_rp}

We give a review on {\em linear random projections},
including {\em normal} and {\em Cauchy random projections}.

Denote the original data matrix by $\mathbf{A} \in
\mathbb{R}^{n\times D}$, i.e., $n$ data points in $D$ dimensions. Let
$\{u_i^\text{T}\}_{i=1}^n \in \mathbb{R}^D$ be the $i$th row of $\mathbf{A}$. Let
$\mathbf{R}\in \mathbb{R}^{D\times k}$ be a random matrix whose
entries are i.i.d. samples of some random variable. The projected
data matrix $\mathbf{B} = \mathbf{AR} \in \mathbb{R}^{n\times
  k}$. Denote the  entries of $\mathbf{R}$ by $\{r_{ij}\}_{i=1}^D\
_{j=1}^k$ and let $\{v_i^\text{T}\}_{i=1}^n \in \mathbb{R}^k$ be the
$i$th row of $\mathbf{B}$. Then $v_i = \mathbf{R}^\text{T}u_i$, with entries $v_{i,j} = \mathbf{R}^\text{T}_ju_i$,
i.i.d. $j = 1$ to $k$, where $\mathbf{R}_j$ is the $j$th column of
$\mathbf{R}$.

For simplicity, we focus on the leading two rows, $u_1$ and $u_2$, in
$\mathbf{A}$, and the leading  two rows, 
$v_1$ and $v_2$, in $\mathbf{B}$. Define $\{x_j\}_{j=1}^k$ to be
\begin{align}
x_j = v_{1,j} - v_{2,j} = \sum_{i=1}^D r_{ij} \left(u_{1,i}-u_{2,i}\right),
\hspace{0.5in} j = 1, 2, ..., k
\end{align}

If we sample $r_{ij}$ i.i.d. from a {\em stable distribution}
\citep{Book:Zolotarev_86,Proc:Indyk_FOCS00}, then $x_j$'s are also
i.i.d. samples of the same stable distribution with a different scale
parameter. In the family of stable distributions, normal and
Cauchy are two important special cases. 

\subsection{Normal Random Projections}

When $r_{ij}$ is sampled from the standard normal, i.e., $r_{ij}\sim
N(0,1)$, i.i.d.,  then 
\begin{align}
x_j = v_{1,j} - v_{2,j} =\sum_{i=1}^D r_{ij} \left(u_{1,i}-u_{2,i}\right) \sim
N\left(0,\sum_{i=1}^D|u_{1,i}-u_{2,i}|^2\right), \ \ \  j = 1, 2, ...,
k, 
\end{align}
\noindent because a weighted sum of normals is also normal. 

Denote the squared $l_2$ distance between $u_1$ and $u_2$ by $d_{l_2} =
\|u_1-u_2\|^2_2 = \sum_{i=1}^D|u_{1,i}-u_{2,i}|^2$. We can estimate
$d_{l_2}$ from the sample squared $l_2$ distance:
\begin{align}
\hat{d}_{l_2} = \frac{1}{k} \sum_{j=1}^k x_j^2.
\end{align}
It is easy to show that (e.g., \citep{Book:Vempala,Proc:Li_Hastie_Church_COLT06})
\begin{align}
&\text{E}\left(\hat{d}_{l_2}\right) = d_{l_2}, \hspace{0.45in}
\text{Var}\left(\hat{d}_{l_2}\right) = \frac{2}{k}d^2_{l_2},\\
&\mathbf{Pr}\left(\left|\hat{d}_{l_2} -d_{l_2}\right|\geq \epsilon d_{l_2}\right)  \leq
2\exp\left(-\frac{k}{4}\epsilon^2 + \frac{k}{6}\epsilon^3\right), \ \
\ \epsilon >0 \label{eqn_normal_tail}
\end{align}

We would like to bound the error probability
$\mathbf{Pr}\left(\left|\hat{d}_{l_2} -d_{l_2}\right|\geq \epsilon
  d_{l_2}\right)$ by $\delta$. Since there
are in total $\frac{n(n-1)}{2} < \frac{n^2}{2}$ pairs among $n$
data points, we need to bound the tail probabilities simultaneously for
all pairs. By the Bonferroni union bound, it suffices if 
\begin{align}
&\frac{n^2}{2}\mathbf{Pr}\left(\left|\hat{d}_{l_2} -d_{l_2}\right|\geq
  \epsilon d_{l_2}\right)  \leq \delta.
\end{align}

Using (\ref{eqn_normal_tail}), it suffices if 
\begin{align}
\frac{n^2}{2}
&2\exp\left(-\frac{k}{4}\epsilon^2 + \frac{k}{6}\epsilon^3\right) \leq
\delta \\
\Longrightarrow & k \geq \frac{2\log n - \log \delta }{\epsilon^2/4 -
  \epsilon^3/6}. 
\end{align}

Therefore, we obtain one version of the JL lemma: 

{\em 
If $k \geq \frac{2\log n - \log \delta }{\epsilon^2/4 -
  \epsilon^3/6}$, then with probability at least $1-\delta$, the
squared $l_2$
distance between any pair of data points (among $n$ data points) can
be approximated within $1\pm \epsilon$ fraction of the
truth, using the squared $l_2$ distance of the
projected data after normal random projections. }

Many versions of the JL lemma have been proved
\citep{Article:JL84,Article:Frankl_JL,Proc:Indyk_STOC98,Proc:Arriaga_FOCS99,Article:Dasgupta_JL,Proc:Indyk_FOCS00,Proc:Indyk_FOCS01,Article:Achlioptas_JCSS03,Article:Proc:Arriaga_Vempala_ML06,Proc:Ailon_STOC06}.

Note that we do not have to use $r_{ij} \sim N(0,1)$ for dimension
reduction in $l_2$. For example, we can sample $r_{ij}$ from
some {\em sub-Gaussian distributions} \citep{Article:Indyk_Naor}, in particular, the following
{\em sparse projection distribution}: 
\begin{align}\label{eqn_subg_rji}
r_{ij} = \sqrt{s}\left\{\begin{array}{rl} 1 & \text{ with prob. }
    \frac{1}{2s}  \\ 0 & \text{ with prob. } 1-\frac{1}{s}\\ -1 & \text{ with prob. }
    \frac{1}{2s} \end{array} \right..
\end{align}

When $ 1\leq s\leq3$, \cite{Article:Achlioptas_JCSS03} proved the JL
lemma for the above sparse
projection, which can also be shown by sub-Gaussian analysis
\citep{Report:Li_Hastie_Church_subrp}. 
Recently,  \cite{Proc:Li_Hastie_Church_KDD06} proposed {\em very
  sparse random projections} using $s = \sqrt{D}$ in
(\ref{eqn_subg_rji}), based on two practical considerations:
\begin{itemize}
\item $D$ should be very large, otherwise
there would be no need for dimension reduction. 
\item 
The original $l_2$ distance should make
engineering sense, in that  the second (or higher) moments should be
bounded (otherwise various {\em term-weighting} schemes will be
applied). 
\end{itemize}

Based on these two practical
assumptions, the projected data are asymptotically normal at a fast
rate of convergence when $s = \sqrt{D}$.  Of course, {\em very sparse
  random projections} do not have worst case performance
guarantees.

\subsection{Cauchy Random Projections}\label{sec_intro}

In {\em Cauchy random projections}, we sample $r_{ij}$ i.i.d. from the
standard Cauchy distribution, i.e., $r_{ij} \sim C(0,1)$. By the 1-stability of Cauchy \citep{Book:Zolotarev_86}, we know that 
\begin{align}
x_j = v_{1,j} - v_{2,j}  \sim C\left(0,\sum_{i=1}^D|u_{1,i} -
  u_{2,i}|\right). 
\end{align}
\noindent That is, the projected differences $x_j = v_{1,j} - v_{2,j}$ are also
Cauchy random variables with the scale parameter being the $l_1$
distance, $d = |u_1 - u_2| = \sum_{i=1}^D|u_{1,i} -
  u_{2,i}|$, in the original space. 

Recall that a Cauchy random variable $z \sim C(0,\gamma)$ has the density 
\begin{align}
f(z)  = \frac{\gamma}{\pi} \frac{1}{z^2 + \gamma^2}, \hspace{0.5in}
\gamma >0, \hspace{0.2in}  -\infty<z<\infty
\end{align}

The easiest way to see the 1-stability is via the characteristic
function, 
\begin{align}
&\text{E}\left(\exp(\sqrt{-1}z_1t)\right) =
\exp\left(-\gamma|t|\right),\\
&\text{E}\left(\exp\left(\sqrt{-1} t\sum_{i=1}^D c_i z_i\right)\right)
= \exp\left(-\gamma\sum_{i=1}^D|c_i|t\right), 
\end{align}
\noindent for $z_1$, $z_2$, ..., $z_D$, i.i.d. $C(0,\gamma)$, and
any constants $c_1$, $c_2$, ..., $c_D$.

Therefore, in {\em Cauchy random projections}, the problem boils down to
estimating the Cauchy scale parameter of $C(0,d)$ from $k$
i.i.d. samples $x_j \sim C(0,d)$.  Unfortunately, unlike in {\em normal
  random projections}, we can no longer estimate $d$ from the
sample mean (i.e., $\frac{1}{k}\sum_{j=1}^k|x_j|$) because
$\text{E}\left(x_j\right) = \infty$.  

Although the impossibility results
\citep{Article:Lee_Naor_04,Article:Brinkman_JACM05}
have ruled out estimators that are metrics, there is enough information
to recover $d$ from $k$ 
samples $\{x_j\}_{j=1}^k$, with a high accuracy.  For
example, \cite{Proc:Indyk_FOCS00} proposed using the sample median as
an estimator. The problem with the sample median estimator is the
inaccuracy at small $k$ and the difficulty in deriving explicit tail
bounds needed for determining the sample size $k$. \\

This study focuses on deriving better estimators and explicit tail bounds for
{\em Cauchy random projections}. Our main results are summarized in
the next section, before we present the detailed derivations. Casual 
readers may skip these derivations after Section
\ref{sec_results}. 

\section{Main Results}\label{sec_results}

 We propose three types of nonlinear
estimators: the bias-corrected sample median estimator
($\hat{d}_{me,c}$), the bias-corrected geometric mean estimator
($\hat{d}_{gm,c}$), and  the bias-corrected maximum likelihood
estimator ($\hat{d}_{MLE,c}$). $\hat{d}_{me,c}$ and $\hat{d}_{gm,c}$
are asymptotically equivalent but the latter is more accurate at small
sample size $k$. In addition, we derive explicit tail bounds for
$\hat{d}_{gm,c}$, from which an analog of the Johnson-Lindenstrauss  (JL)
lemma for dimension reduction in $l_1$ follows. Asymptotically, both
$\hat{d}_{me,c}$ and $\hat{d}_{gm,c}$ are $\frac{8}{\pi^2} \approx
80\%$ efficient compared to the maximum likelihood estimator
$\hat{d}_{MLE,c}$. We propose accurate approximations to the
distribution and tail bounds of $\hat{d}_{MLE,c}$, while the exact
closed-form answers are not attainable. 

\subsection{The Bias-corrected Sample Median Estimator}

Denoted by $\hat{d}_{me,c}$, the bias-corrected sample median
estimator is
\begin{align}
\hat{d}_{me,c} = \frac{\hat{d}_{me}}{b_{me}}, 
\end{align}
\noindent where
\begin{align}
\hat{d}_{me} &= \text{median}(|x_j|, j = 1, 2,..., k)\\
b_{me}
&=
\int_0^1\frac{(2m+1)!}{(m!)^2}\tan\left(\frac{\pi}{2}t\right)\left(t-t^2\right)^m
  dt, \ \ \ k = 2m+1 
\end{align}

Here, for convenience, we only consider $k = 2m+1$, $m$ = 1, 2, 3,
...

Some key properties of $\hat{d}_{me,c}$: 

\begin{itemize}
\item $\text{E}\left(\hat{d}_{me,c}\right) = d$, i.e, $\hat{d}_{me,c}$
  is unbiased. 
\item When $k\geq 5$, the variance of $\hat{d}_{me,c}$ is 
\begin{align}
\text{Var}\left(\hat{d}_{me,c}\right) =
d^2\left(\frac{(m!)^2}{(2m+1)!}\frac{\int_0^1
  \tan^2\left(\frac{\pi}{2}t\right)\left(t-t^2\right)^m dt}{\left(\int_0^1
  \tan\left(\frac{\pi}{2}t\right)\left(t-t^2\right)^m dt\right)^2} -
1\right), \ \ \ \ k\geq5
\end{align}
$\text{Var}\left(\hat{d}_{me,c}\right) = \infty$ if $k = 3$. 
\item As $k \rightarrow \infty$, $\hat{d}_{me,c}$ converges to a
  normal in distribution 
\begin{align}
\sqrt{k}\left(\hat{d}_{me,c}  - d \right)\overset{D}{\Longrightarrow} N\left(0,\frac{\pi^2}{4}d^2\right).
\end{align}
\end{itemize}

\subsection{The Bias-corrected Geometric Mean Estimator}
Denoted by $\hat{d}_{gm,c}$, the bias-corrected geometric mean
estimator is defined as 
\begin{align}
\hat{d}_{gm,c} =
\cos^k\left(\frac{\pi}{2k}\right)\prod_{j=1}^k|x_j|^{1/k},
\hspace{0.1in} k>1
\end{align}

Important properties of $\hat{d}_{gm,c}$ include: 
\begin{itemize}
\item This estimator is a non-convex norm, i.e., the $l_p$ norm
  with $p\rightarrow 0$. 
\item It is unbiased, i.e., $\text{E}\left(\hat{d}_{gm,c}\right)
  = d$. 
\item Its variance is (for $k>2$) 
\begin{align}
\text{Var}\left(\hat{d}_{gm,c}\right) &= d^2
\left(\frac{\cos^{2k}\left(\frac{\pi}{2k}\right)}{\cos^k\left(\frac{\pi}{k}\right)}-1
\right)
= \frac{\pi^2}{4}\frac{d^2}{k} +
\frac{\pi^4}{32}\frac{d^2}{k^2}+O\left(\frac{1}{k^3}\right).
\end{align}
\item For $0\leq \epsilon \leq 1$, its tail bounds can be represented in exponential forms 
\begin{align}
&\mathbf{Pr}\left(\hat{d}_{gm,c} - d > \epsilon d \right) \leq
\exp\left(-k\left(\frac{\epsilon^2}{8(1+\epsilon)}\right)\right)\\
&\mathbf{Pr}\left(\hat{d}_{gm,c} - d < -\epsilon d \right) \leq
\exp\left(-k\left(\frac{\epsilon^2}{8(1+\epsilon)}\right)\right), \ \
\ k \geq \frac{\pi^2}{1.5\epsilon}
\end{align}
\item These exponential tail bounds yield an analog of the 
Johnson-Lindenstrauss  (JL) lemma for dimension reduction in $l_1$:

{\em 
If $k \geq \frac{8\left(2\log n -
  \log\delta\right)}{\epsilon^2/(1+\epsilon)}\geq \frac{\pi^2}{1.5\epsilon}$, then with probability at
least $1-\delta$, one can recover the original $l_1$ distance between
any pair of data points (among all $n$ data points) within 
$1\pm\epsilon$ ($0\leq
\epsilon\leq 1$) fraction of the truth,
using $\hat{d}_{gm,c}$, i.e., $|\hat{d}_{gm,c}-d|\leq \epsilon d$. }
\end{itemize}

\subsection{The Bias-corrected Maximum Likelihood Estimator} 
Denoted by $\hat{d}_{MLE,c}$, the bias-corrected maximum likelihood
estimator is 
\begin{align}
\hat{d}_{MLE,c} = \hat{d}_{MLE}\left(1-\frac{1}{k}\right),
\end{align}
where $\hat{d}_{MLE}$ solves a nonlinear MLE equation 
\begin{align}
-\frac{k}{\hat{d}_{MLE}} + \sum_{j=1}^k\frac{2\hat{d}_{MLE}}{x_j^2 + \hat{d}_{MLE}^2} = 0.
\end{align}

Some properties of $\hat{d}_{MLE,c}$:
\begin{itemize}
\item It is nearly unbiased, $\text{E}\left(\hat{d}_{MLE,c}\right) = d
  + O\left(\frac{1}{k^2}\right)$. 
\item Its asymptotic variance is 
\begin{align}
\text{Var}\left(\hat{d}_{MLE,c}\right) = \frac{2d^2}{k} +
\frac{3d^2}{k^2} 
  + O\left(\frac{1}{k^3}\right), 
\end{align}
\noindent i.e.,
$\frac{\text{Var}\left(\hat{d}_{MLE,c}\right)}{\text{Var}\left(\hat{d}_{me,c}\right)}
\rightarrow \frac{8}{\pi^2}$, $\frac{\text{Var}\left(\hat{d}_{MLE,c}\right)}{\text{Var}\left(\hat{d}_{gm,c}\right)}
\rightarrow \frac{8}{\pi^2}$, as $k\rightarrow
\infty$. ($\frac{8}{\pi^2} \approx 80\%$) 
\item Its distribution can be accurately approximated by an inverse
  Gaussian, at least in the small deviation range. Based on the
  inverse Gaussian approximation, we suggest the following approximate tail bound
\begin{align}
&\mathbf{Pr}\left(|\hat{d}_{MLE,c} - d| \geq \epsilon d\right) \overset{\sim}{\leq}
2\exp\left(-\frac{\epsilon^2/(1+\epsilon)}{2 \left(\frac{2}{k} + \frac{3}{k^2}\right)}\right),
\hspace{0.15in} 0\leq \epsilon \leq 1, 
\end{align}
\noindent which has been verified by simulations for the tail
probability $\geq 10^{-10}$ range. 
\end{itemize}

\section{The Sample Median Estimators}\label{sec_median}

Recall in Cauchy random projections, $\mathbf{B} = \mathbf{AR}$, we
denote the leading two rows in $\mathbf{A}$ by $u_1$, $u_2$ $\in
\mathbb{R}^{D}$, and the leading two rows in $\mathbf{B}$ by $v_1$,
$v_2$ $\in \mathbb{R}^{k}$. Our goal is to estimate the $l_1$ distance
$d = |u_1 - u_2| = \sum_{i=1}^D |u_{1,i} - u_{2,i}|$ from
$\{x_j\}_{j=1}^k$, $x_j = v_{1,j} - v_{2,j} \sim C(0,d)$, i.i.d.

It is easy to show (e.g., \cite{Proc:Indyk_FOCS00}) that the
population median of $|x_j|$ is $d$. Therefore, it is natural to
consider estimating $d$ from the sample median,
\begin{align} \label{eqn_def_me}
\hat{d}_{me} = \text{median}\{|x_j|, j = 1, 2, ..., k\}.
\end{align}

As illustrated in the following lemma (proved in Appendix \ref{app_proof_lem_me}), the sample median estimator,
$\hat{d}_{me}$, is asymptotically  unbiased and normal. For small
samples (e.g., $k\leq 20$), however, $\hat{d}_{me}$ is severely
biased. 

\begin{lemma} \label{lem_me}
The sample median estimator, $\hat{d}_{me}$, defined in
(\ref{eqn_def_me}), is asymptotically unbiased and normal 
\begin{align}
\sqrt{k}\left(\hat{d}_{me}  - d \right)\overset{D}{\Longrightarrow} N\left(0,\frac{\pi^2}{4}d^2\right)
\end{align}
When $k = 2m+1$, $m$ = 1, 2, 3, ..., the $r^{th}$ moment of
$\hat{d}_{me}$ can be represented as 
\begin{align}
&\text{E}\left(\hat{d}_{me}\right)^r = d^r\left(\int_0^1\frac{(2m+1)!}{(m!)^2}\tan^r\left(\frac{\pi}{2}t\right)\left(t-t^2\right)^m
  dt\right), \ \ \  m \geq r
\end{align}
If $m<r$, then $\text{E}\left(\hat{d}_{me}\right)^r = \infty$. \\ \\
\end{lemma}

For simplicity, we only consider $k = 2m+1$ when evaluating 
$\text{E}\left(\hat{d}_{me}\right)^r$. 

Once we know $\text{E}\left(\hat{d}_{me}\right)$, we can remove the
bias of $\hat{d}_{me}$ using 
\begin{align}
\hat{d}_{me,c} = \frac{\hat{d}_{me}}{b_{me}},
\end{align}
where the bias correction factor $b_{me}$ is 
\begin{align}\label{eqn_bme}
b_{me} = \frac{\text{E}\left(\hat{d}_{me}\right)}{d} = \int_0^1\frac{(2m+1)!}{(m!)^2}\tan\left(\frac{\pi}{2}t\right)\left(t-t^2\right)^m
  dt.
\end{align}

$b_{me}$ can be numerically evaluated and tabulated, at least for small
$k$.\footnote{It is possible to express $b_{me}$ as an infinite
  sum. Note that $\frac{(2m+1)!}{(m!)^2}\left(t-t^2\right)^m$, $0\leq
  t\leq 1$, is the probability density of a Beta distribution
  $Beta(m+1,m+1)$.}
% By Taylor expansion \citep[1.411.6]{Book:Gradshteyn_94},
%  $\tan\left(\frac{\pi}{2}t\right) =
%  \sum_{j=1}^\infty\frac{2^{2j}\left(2^{2j}-1\right)}{(2j)!}|B_{2j}|\left(\frac{\pi}{2}\right)^{2j-1}t^{2j-1}$, where $B_{2j}$ is the {\em Bernoulli number} \citep[9.61]{Book:Gradshteyn_94}. If $z \sim Beta(m+1,m+1)$, then $\text{E}\left(z^r\right) = \frac{(2m+1)!(m+r)!}{(2m+1+r)!m!}$ (\url{http://mathworld.wolfram.com/BetaDistribution.html}). Therefore, $b_{me} = \sum_{j=1}^\infty\frac{2^{2j}\left(2^{2j}-1\right)}{(2j)!}|B_{2j}|\left(\frac{\pi}{2}\right)^{2j-1} \frac{(2m+1)!(m+2j-1)!}{(2m+2j)!m!}$. }

Obviously, $\hat{d}_{me,c}$ is unbiased, i.e.,
$\text{E}\left(\hat{d}_{me,c}\right) = d$. Its variance would be 
\begin{align}
\text{Var}\left(\hat{d}_{me,c}\right) =
d^2\left(\frac{(m!)^2}{(2m+1)!}\frac{\int_0^1
  \tan^2\left(\frac{\pi}{2}t\right)\left(t-t^2\right)^m dt}{\left(\int_0^1
  \tan\left(\frac{\pi}{2}t\right)\left(t-t^2\right)^m dt\right)^2} -
1\right), \ \ \ \ k=2m+1\geq5
\end{align}

Of course, $\hat{d}_{gm,c}$ and $\hat{d}_{gm}$ are asymptotically
equivalent, i.e., 
$\sqrt{k}\left(\hat{d}_{me,c}  - d \right)\overset{D}{\Longrightarrow}
N\left(0,\frac{\pi^2}{4}d^2\right)$. 

\begin{figure}[h]
\begin{center}
\includegraphics[width = 2.5in]{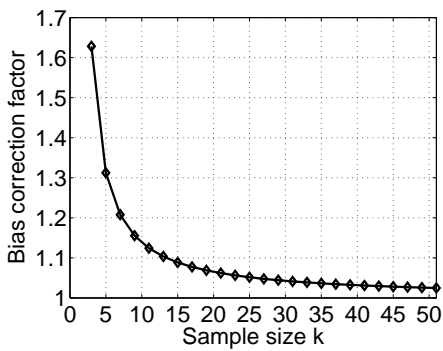}
\end{center}\vspace{-0.3in}
\caption{The bias correction factor, $b_{me}$ in (\ref{eqn_bme}), as a function of $k
  =2m+1$. After $k>50$, the bias is negligible. Note that
  $b_{me}=\infty$ when $k=1$. }\label{fig_bme}
\end{figure}

Figure \ref{fig_bme} plots $b_{me}$ as a function of $k$, indicating
that $\hat{d}_{me}$ is severely biased when $k\leq 20$. When $k>50$,
the bias becomes negligible. Note that, because $b_{me}\geq 1$, the bias
correction not only removes the bias of $\hat{d}_{me}$ but also
reduces its variance.

The sample median is a special case of sample quantile estimators
\citep{Article:Fama_68,Article:Fama_71}.   For example, one 
version of the quantile estimators given by
\cite{Article:McCulloch_86} would be
\begin{align}
\hat{d}_{or} = \frac{\hat{|x|}_{.75} - \hat{|x|}_{.25}}{2.0},
\end{align}
\noindent where $\hat{|x|}_{.75}$ and $\hat{|x|}_{.25}$ are the .75 and
.25 sample quantiles of $\{|x_{j}|\}_{j=1}^k$, respectively. 

Our simulations indicate that $\hat{d}_{me}$ actually slightly outperforms
$\hat{d}_{or}$. This is not surprising. $\hat{d}_{or}$ works for any
Cauchy distribution whose location parameter does not have to be zero,
while $\hat{d}_{me}$ takes advantage of the fact that the
Cauchy location parameter is always zero in our case.

\section{The Geometric Mean Estimators }\label{sec_gm}

This section derives estimators based on the geometric
mean, which are more accurate than the sample median estimators. The
geometric mean estimators allow us to derive tail bounds in explicit
forms and (consequently) an analog of the
Johnson-Lindenstrauss  (JL) lemma for dimension reduction in $l_1$. 

Recall, our goal is to estimate $d$ from $k$ i.i.d. samples $x_j
\sim C(0,d)$. To help derive the geometric mean estimators, we
first study two nonlinear estimators based on the fractional moment, i.e., $\text{E}(|x|^\lambda)$
($|\lambda|<1$) and the logarithmic moment, i.e,
$\text{E}\left(\log(|x|)\right)$, respectively, as presented in 
Lemma  \ref{lem_d_log}. See the proof in Appendix \ref{app_proof_lem_d_log}. 

\begin{lemma}\label{lem_d_log}
Assume $x \sim C(0,d)$. Then
\begin{align}
&\text{E}\left(|x|^\lambda\right) 
=\frac{d^\lambda}{\cos(\lambda\pi/2)}, \hspace{0.5in}|\lambda|<1\\
&\text{E}\left(\log(|x|)\right) = \log(d), \\
&\text{Var}\left(\log(|x|)\right) = \frac{\pi^2}{4}, 
\end{align}
\noindent from which we can derive two biased estimators of $d$ from
$k$ i.i.d. samples $x_j \sim C(0,d)$:
\begin{align}
&\hat{d}_\lambda = \left(\frac{1}{k}\sum_{j=1}^k|x_j|^\lambda
  \cos(\lambda\pi/2)\right)^{1/\lambda}, \hspace{0.2in} |\lambda| <1,\\
&\hat{d}_{log} = \exp\left(\frac{1}{k}\sum_{j=1}^k\log(|x_j|)\right),
\end{align}
\noindent whose variances are, respectively,
\begin{align}
&\text{Var}\left(\hat{d}_{\lambda}\right) = \frac{d^2}{k}
\frac{\sin^2(\lambda \pi/2)}{\lambda^2 \cos(\lambda\pi)} +
O\left(\frac{1}{k^2}\right), \hspace{0.2in} |\lambda| <1/2\\
&\text{Var}\left(\hat{d}_{log}\right)  = \frac{\pi^2d^2}{4k} +
O\left(\frac{1}{k^2}\right).
\end{align}

The term $\frac{\sin^2(\lambda \pi/2)}{\lambda^2 \cos(\lambda\pi)}$
decreases with decreasing $|\lambda|$, reaching a limit
\begin{align}
\underset{\lambda\rightarrow 0}\lim\frac{\sin^2(\lambda
  \pi/2)}{\lambda^2 \cos(\lambda\pi)} = \frac{\pi^2}{4}.
\end{align}
\noindent In other words, the variance of $\hat{d}_{\lambda}$ converges to
that of $\hat{d}_{log}$ as $|\lambda|$ approaches zero. 
\\
\end{lemma}

 Note that $\hat{d}_{log}$ can in fact be
written as the {\em geometric mean}:
\begin{align}
\hat{d}_{log} = \hat{d}_{gm} = \prod_{j=1}^k|x_j|^{1/k}. 
\end{align}

$\hat{d}_{\lambda}$ is a non-convex norm ($l_\lambda$) because $\lambda
<1$. $\hat{d}_{gm}$ is also
a non-convex norm (the $l_\lambda$ norm as $\lambda \rightarrow 0$). Both
$\hat{d}_{\lambda}$ and $\hat{d}_{gm}$ do not satisfy the triangle
inequality. 

We propose $\hat{d}_{gm,c}$, the bias-corrected geometric mean
estimator. Lemma \ref{lem_d_gm}  derives the moments of
$\hat{d}_{gm,c}$, proved in Appendix \ref{app_proof_lem_d_gm}.

\begin{lemma}\label{lem_d_gm}
\begin{align}
\hat{d}_{gm,c} =
\cos^k\left(\frac{\pi}{2k}\right)\prod_{j=1}^k|x_j|^{1/k},
\hspace{0.1in} k>1
\end{align}
is unbiased, with the variance  (valid when $k>2$)
\begin{align}
\text{Var}\left(\hat{d}_{gm,c}\right) &= d^2
\left(\frac{\cos^{2k}\left(\frac{\pi}{2k}\right)}{\cos^k\left(\frac{\pi}{k}\right)}-1
\right)=\frac{d^2}{k} \frac{\pi^2}{4} +
\frac{\pi^4}{32}\frac{d^2}{k^2}+O\left(\frac{1}{k^3}\right).
\end{align}

The third and fourth central moments are  (for $k>3$ and $k>4$,
respectively) 
\begin{align}
&\text{E}\left(\hat{d}_{gm,c} -
  \text{E}\left(\hat{d}_{gm,c}\right)\right)^3 =
\frac{3\pi^4}{16}\frac{d^3}{k^2} + O\left(\frac{1}{k^3}\right) \\
&\text{E}\left(\hat{d}_{gm,c} -
  \text{E}\left(\hat{d}_{gm,c}\right)\right)^4 =
\frac{3\pi^4}{16}\frac{d^4}{k^2} + O\left(\frac{1}{k^3}\right).
\end{align}\\
\end{lemma}

The higher (third or fourth) moments may be useful for approximating
the distribution of $\hat{d}_{gm,c}$.  In Section \ref{sec_mle}, we
will show how to approximate the distribution of the maximum
likelihood estimator by matching the first four moments (in the
leading terms). We could apply the similar technique to approximate
$\hat{d}_{gm,c}$. Fortunately, we do not have to do so because we are
able to derive the exact tail bounds of $\hat{d}_{gm,c}$ in Lemma
\ref{lem_d_gm_tail}, which is proved in Appendix \ref{app_proof_lem_d_gm_tail}.

\begin{lemma}\label{lem_d_gm_tail}
\begin{align}\label{eqn_gm_bound}
\mathbf{Pr}\left(\hat{d}_{gm,c} \geq (1+\epsilon)d \right) \leq
\frac{\cos^{kt_1^*}\left(\frac{\pi}{2k}\right)}{\cos^k\left(\frac{\pi
      t_1^*}{2k}\right)(1+\epsilon)^{t_1^*}}, \hspace{0.25in} \epsilon \geq0
\end{align}
\noindent where 
\begin{align}
t_1^* = \frac{2k}{\pi}\tan^{-1}\left(\left(\log(1+\epsilon) -
    k\log\cos\left(\frac{\pi}{2k}\right)\right)\frac{2}{\pi}\right). 
\end{align}
\begin{align}\label{eqn_gm_bound_left}
\mathbf{Pr}\left(\hat{d}_{gm,c} \leq  (1-\epsilon)d \right) \leq
\frac{ (1-\epsilon)^{t_2^*}}{\cos^k\left(\frac{\pi
      t_2^*}{2k}\right)\cos^{kt_2^*}\left(\frac{\pi}{2k}\right)},
\hspace{0.25in} 0\leq \epsilon\leq 1, \hspace{0.1in} k\geq \frac{\pi^2}{8\epsilon}
\end{align}
\noindent where 
\begin{align}
t_2^* = \frac{2k}{\pi}\tan^{-1}\left(\left(-\log(1-\epsilon) +
    k\log\cos\left(\frac{\pi}{2k}\right)\right)\frac{2}{\pi}\right). 
\end{align}

By restricting $0\leq\epsilon\leq 1$, the tail bounds can be written
in exponential forms: 
\begin{align}\label{eqn_exp_right}
&\mathbf{Pr}\left(\hat{d}_{gm,c} \geq (1+\epsilon)d \right) \leq
\exp\left(-k\frac{\epsilon^2}{8(1+\epsilon)}\right) \\
&\mathbf{Pr}\left(\hat{d}_{gm,c} \leq (1-\epsilon)d \right) \leq
\exp\left(-k\frac{\epsilon^2}{8(1+\epsilon)}\right), \hspace{0.2in} k\geq \frac{\pi^2}{1.5\epsilon}\label{eqn_exp_left}
\end{align}\\
\end{lemma}

An analog of the JL bound for $l_1$ follows from the exponential tail
bounds (\ref{eqn_exp_right}) and 
(\ref{eqn_exp_left}). 
\begin{lemma}\label{lem_JL_l1}
Using $\hat{d}_{gm,c}$ with $k \geq \frac{8\left(2\log n -
  \log\delta\right)}{\epsilon^2/(1+\epsilon)} \geq
\frac{\pi^2}{1.5\epsilon}$, then with probability at
least $1-\delta$, the $l_1$ distance, $d$, between
any pair of data points (among $n$ data points), can be estimated with
errors bounded by $\pm \epsilon d$, i.e., $|\hat{d}_{gm,c} - d| \leq
\epsilon d$.  
\end{lemma}

\textbf{Remarks on Lemma \ref{lem_JL_l1}}: (1) We can replace the constant ``8'' in Lemma
\ref{lem_JL_l1} with better (i.e., smaller) constants for 
specific values of $\epsilon$. For example, If $\epsilon = 0.2$, we can
replace ``8'' by ``5''. See the proof of Lemma \ref{lem_d_gm_tail}. 
(2) This Lemma is weaker than the classical JL Lemma for
dimension reduction in $l_2$ as reviewed in Section 2.1. The classical
JL Lemma for $l_2$ ensures that the $l_2$ inter-point distances of the
projected data points are close enough to the original $l_2$
distances, while Lemma
\ref{lem_JL_l1} merely says that the projected data points contain
enough information to reconstruct the original $l_1$ distances.  On
the other hand, the geometric mean estimator is a non-convex
norm; and therefore it does contain some information about the
geometry. We leave it for future work to explore the possibility of
developing efficient algorithms using the geometric mean estimator. \\

\begin{figure}[h]
\begin{center}\mbox{
\subfigure[$k=5$]{\includegraphics[width = 2.5in]{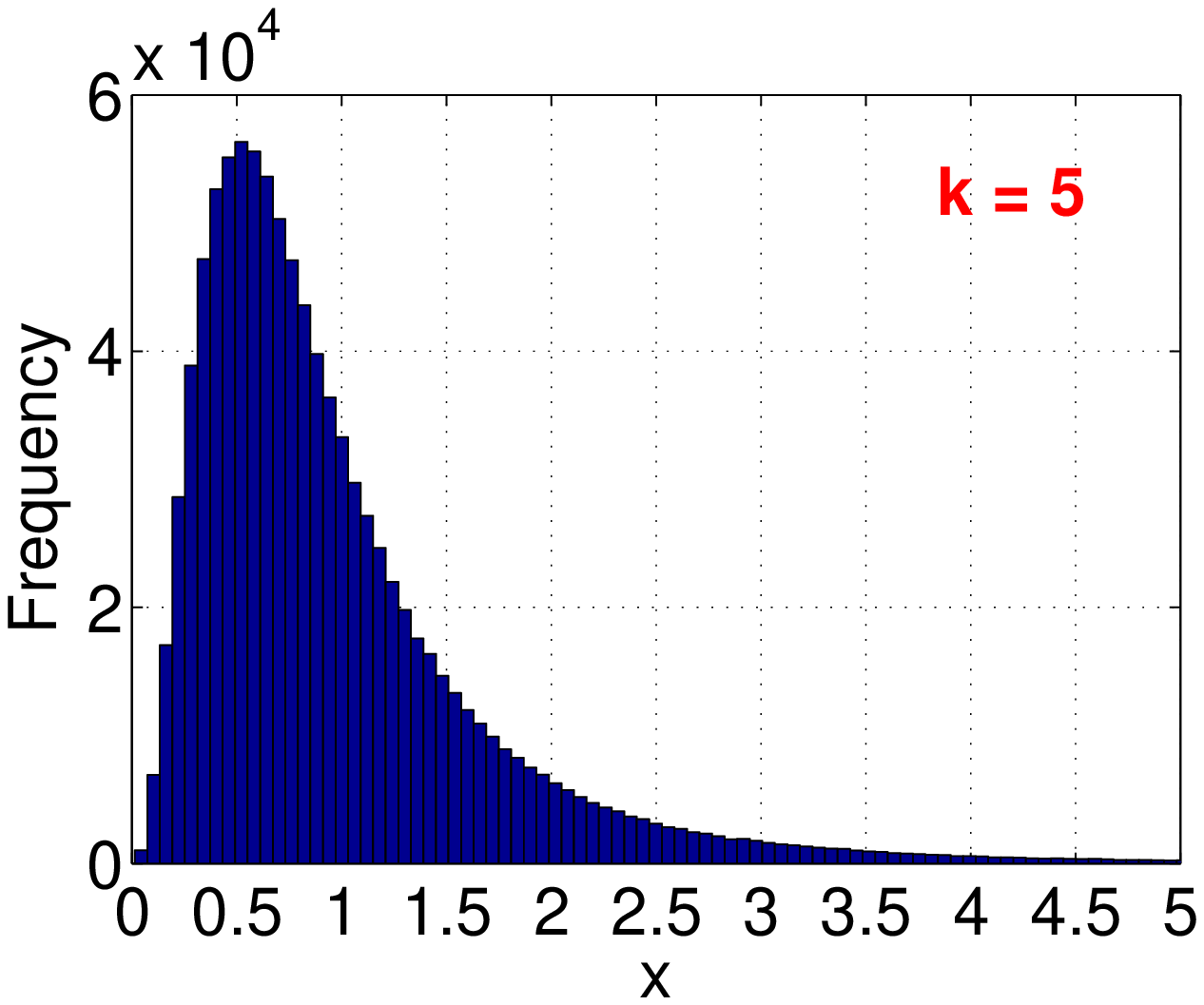}}
\subfigure[$k=50$]{\includegraphics[width = 2.5in]{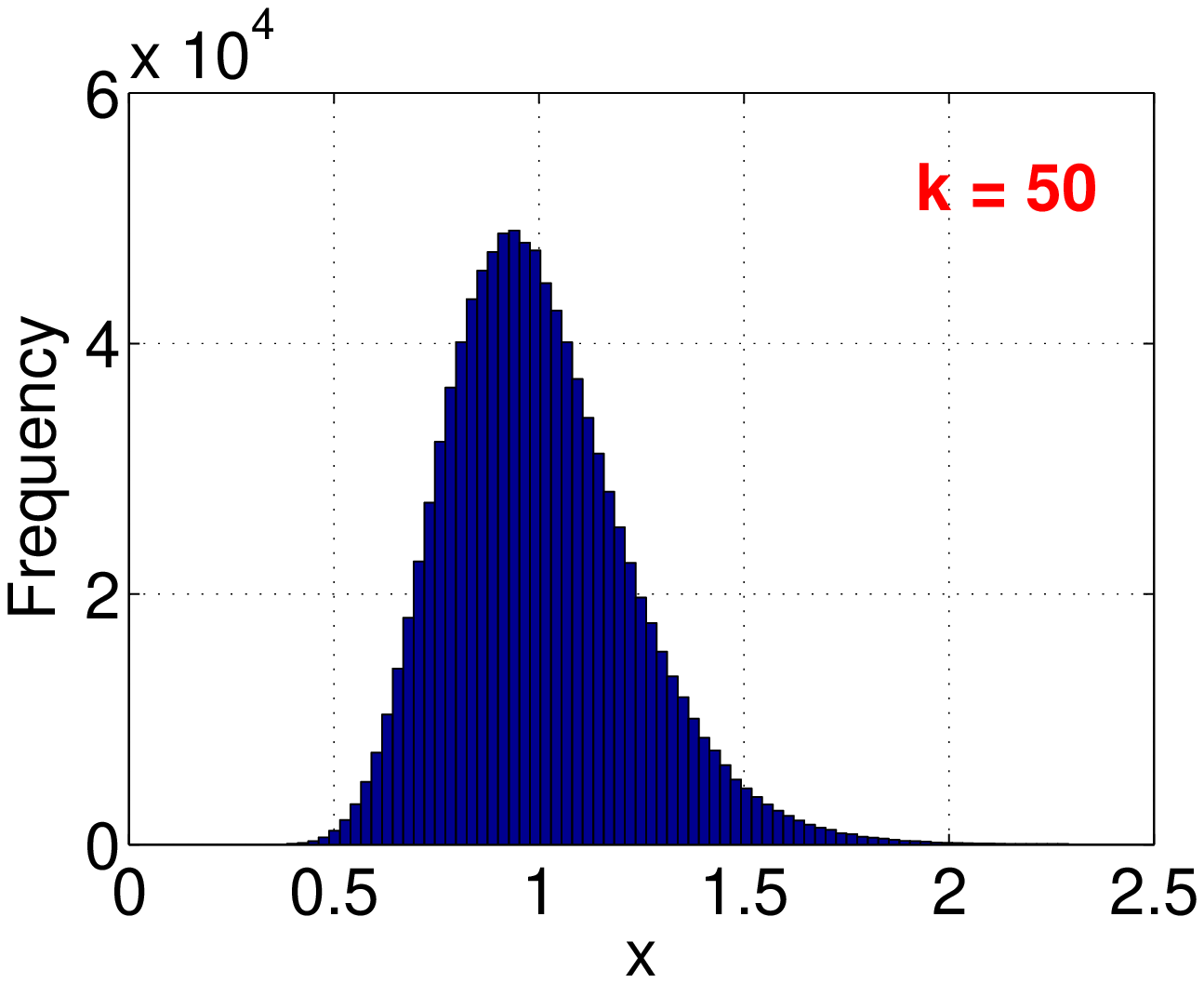}}}
\end{center}\vspace{-0.4in}
\caption{Histograms of $\hat{d}_{gm,c}$, obtained from $10^6$ simulations. At
  least in the range not too far from the mean, the
  distribution of $\hat{d}_{gm,c}$ resembles a gamma and also resembles
a normal when $k$ is large enough. }\label{fig_hist_d_gm}
\end{figure}

Figure \ref{fig_hist_d_gm}   presents the simulated histograms of $\hat{d}_{gm,c}$
for $d=1$, with $k = 5$ and $k=50$. The histograms reveal some
characteristics shared by the maximum likelihood estimator  we will
discuss in the next section: 
\begin{itemize}
\item Supported on $[0,\infty)$, $\hat{d}_{gm,c}$ is positively
skewed. 
\item The distribution of $\hat{d}_{gm,c}$ is still
  ``heavy-tailed.'' However, in the region not too far from the mean, the distribution of $\hat{d}_{gm,c}$ may be
  well captured by a gamma (or a generalized gamma) distribution. For large $k$, even a 
normal  approximation may suffice. 
\end{itemize}

\begin{figure}[h]
\begin{center}
\includegraphics[width = 2.5in]{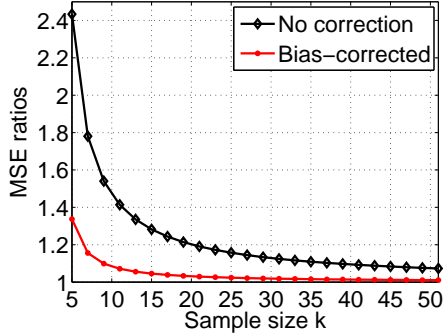}
\end{center}\vspace{-0.25in}
\caption{ The ratios of the mean square errors (MSN),
  $\frac{\text{MSE}(\hat{d}_{me})}{\text{MSE}(\hat{d}_{gm,c})}$ and
  $\frac{\text{MSE}(\hat{d}_{me,c})}{\text{MSE}(\hat{d}_{gm,c})}$,
  demonstrate that the bias-corrected geometric mean estimator
  $\hat{d}_{gm,c}$ is considerably more accurate than the sample
  median estimator $\hat{d}_{me}$. The bias correction on
  $\hat{d}_{me}$ considerably reduces the MSE. Note that when $k=3$, the ratios are $\infty$. }\label{fig_gm_vs_me}
\end{figure}

Figure \ref{fig_gm_vs_me} compares $\hat{d}_{gm,c}$ with the sample median estimators $\hat{d}_{me}$ and
$\hat{d}_{me,c}$, in terms of the mean square errors.  $\hat{d}_{gm,c}$ is considerably more accurate than
$\hat{d}_{me}$ at small $k$. The bias correction significantly reduces
the mean square errors of $\hat{d}_{me}$.

\section{The Maximum Likelihood Estimators}\label{sec_mle}

This section is devoted to analyzing the maximum likelihood
estimators (MLE), which are ``asymptotically optimum.'' In comparisons, 
the sample median estimators and geometric mean estimators are
not optimum.  Our contribution in this section includes the higher-order
analysis for the bias and  moments and accurate closed-from
approximations to the distribution of the MLE.

The method of maximum likelihood is widely used.  For example, \cite{Proc:Li_Hastie_Church_COLT06} applied the maximum likelihood method to {\em normal random
  projections} and provided an improved estimator of the
$l_2$ distance by taking advantage of the marginal information.

The Cauchy distribution is often considered a ``challenging''
example because of the ``multiple
roots'' problem when estimating the location
parameter \citep{Article:Barnett_66,Article:Haas_70}. In our case, since
the location parameter is always zero, much of the difficulty is avoided. 

Recall our goal is to estimate $d$ from $k$ i.i.d. samples
$x_j \sim C(0,d), j = 1, 2,..., k$. The $\log$ joint
likelihood of $\{x_j\}_{j=1}^k$ is  
\begin{align}
L(x_1,x_2,...x_k;d) = k\log(d) - k\log(\pi) - \sum_{j=1}^k\log(x_j^2+d^2),
\end{align}
\noindent whose first and second derivatives (w.r.t. $d$) are
\begin{align}
&L^\prime(d) = \frac{k}{d} - \sum_{j=1}^k\frac{2d}{x_j^2+d^2},\\
&L^{\prime\prime}(d) = -\frac{k}{d^2} -
\sum_{j=1}^k\frac{2x_j^2-2d^2}{(x_j^2+d^2)^2} =
- \frac{ L^\prime(d)}{d}  - 4\sum_{j=1}^k\frac{x_j^2}{(x_j^2+d^2)^2}.
\end{align}

The maximum likelihood estimator of $d$, denoted by $\hat{d}_{MLE}$, is 
the solution  to $L^\prime(d) = 0$, i.e., 
\begin{align}\label{eqn_mle}
-\frac{k}{\hat{d}_{MLE}}+\sum_{j=1}^k\frac{2\hat{d}_{MLE}}{x_j^2+\hat{d}_{MLE}^2} = 0.
\end{align}
\noindent Because $L^{\prime\prime}(\hat{d}_{MLE}) \leq 0$, $\hat{d}_{MLE}$ indeed maximizes the joint likelihood and is the
only solution to the MLE equation (\ref{eqn_mle}). Solving
(\ref{eqn_mle}) numerically is not difficult (e.g., a few iterations
using the Newton's method). For a better accuracy, we
recommend the following bias-corrected estimator:
\begin{align}
\hat{d}_{MLE,c} = \hat{d}_{MLE}\left(1-\frac{1}{k}\right).
\end{align}

Lemma  \ref{lem_mle_asymp} concerns the asymptotic moments of $\hat{d}_{MLE}$ and $\hat{d}_{MLE,c}$, proved in Appendix
\ref{app_proof_lem_asymp}. 
\begin{lemma}\label{lem_mle_asymp}
Both $\hat{d}_{MLE}$ and $\hat{d}_{MLE,c}$ are asymptotically unbiased and
normal. The first four moments of $\hat{d}_{MLE}$ are
\begin{align}
&\text{E}\left(\hat{d}_{MLE} - d\right) = \frac{d}{k}+ O\left(\frac{1}{k^2}\right) \\
&\text{Var}\left(\hat{d}_{MLE}\right) = \frac{2d^2}{k} + \frac{7d^2}{k^2} +O\left(\frac{1}{k^3}\right)\\
&\text{E}\left(\hat{d}_{MLE} - \text{E}(\hat{d}_{MLE})\right)^3 = \frac{12d^3}{k^2} +
O\left(\frac{1}{k^3}\right) \\
&\text{E}\left(\hat{d}_{MLE} - \text{E}(\hat{d}_{MLE})\right)^4 = \frac{12d^4}{k^2} +
\frac{222d^4}{k^3} + O\left(\frac{1}{k^4}\right)
\end{align}
The first four moments of $\hat{d}_{MLE,c}$ are
\begin{align}
&\text{E}\left(\hat{d}_{MLE,c} - d\right) =
O\left(\frac{1}{k^2}\right) \\
&\text{Var}\left(\hat{d}_{MLE,c}\right) = \frac{2d^2}{k} +
\frac{3d^2}{k^2}+O\left(\frac{1}{k^3}\right)  \\
&\text{E}\left(\hat{d}_{MLE,c} - \text{E}(\hat{d}_{MLE,c})\right)^3 = \frac{12d^3}{k^2} +
O\left(\frac{1}{k^3}\right) \\
&\text{E}\left(\hat{d}_{MLE,c} - \text{E}(\hat{d}_{MLE,c})\right)^4 = \frac{12d^4}{k^2} +
\frac{186d^4}{k^3} + O\left(\frac{1}{k^4}\right) 
\end{align}\\
\end{lemma}

The order $O\left(\frac{1}{k}\right)$ term of the
variance, i.e.,  $\frac{2d^2}{k}$, is known, e.g.,
 \citep{Article:Haas_70}.  We derive the  bias-corrected estimator, $\hat{d}_{MLE,c}$,  and the higher order moments using stochastic Taylor
expansions \citep{Article:Bartlett_53,Article:Shenton_63,Article:Ferrari_96,Article:Cysneiros_01}.

We will propose an inverse Gaussian distribution to approximate the
distribution of $\hat{d}_{MLE,c}$, by matching the first four moments
(at least in the leading terms). 

\subsection{A Numerical Example}
\begin{figure}[h]
\begin{center}\mbox{
\subfigure[{\scriptsize $\text{E}(\hat{d}_{MLE}-d)/d$ v.s. $\text{E}(\hat{d}_{MLE,c}-d)/d$}]{\includegraphics[width = 2.25in]{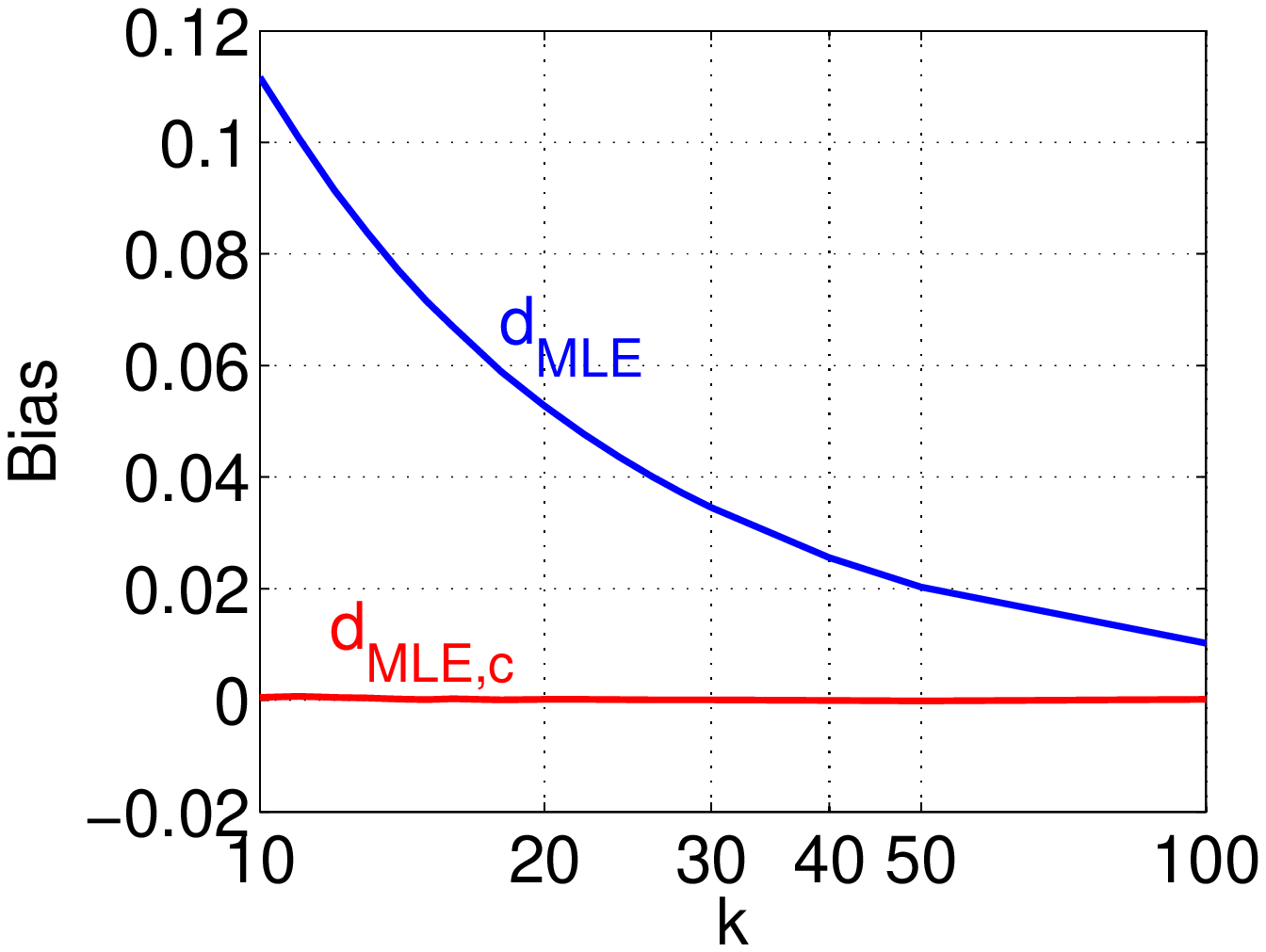}}
\subfigure[{\scriptsize $\left(\text{E}(\hat{d}_{MLE,c}-\text{E}(\hat{d}_{MLE,c}))^2/d^2\right)^{1/2}$}]{\includegraphics[width = 2.25in]{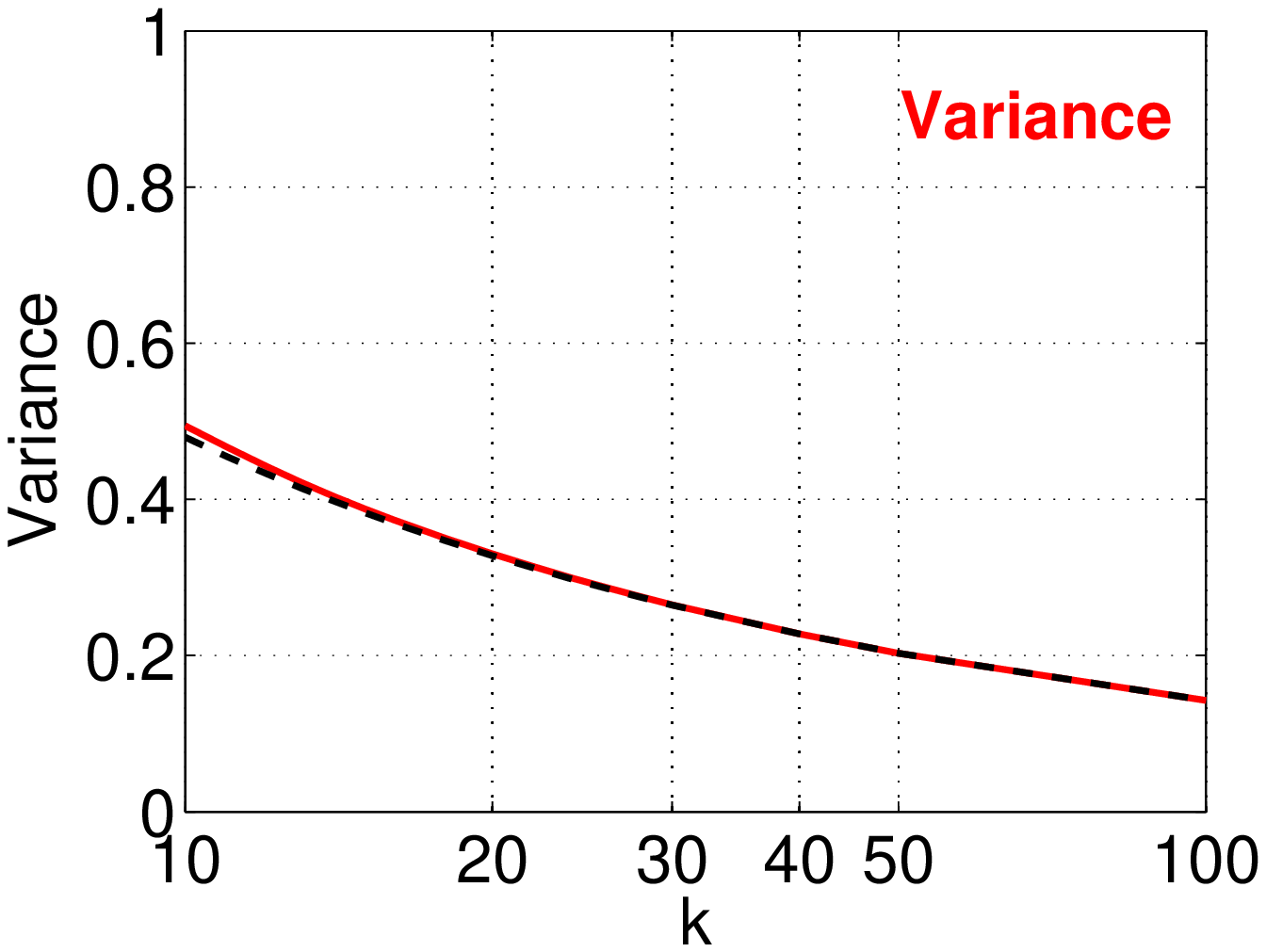}}}\vspace{-0.3in}
\mbox{
\subfigure[{\scriptsize $\left(\text{E}(\hat{d}_{MLE,c}-\text{E}(\hat{d}_{MLE,c}))^3/d^3\right)^{1/3}$}]{\includegraphics[width = 2.25in]{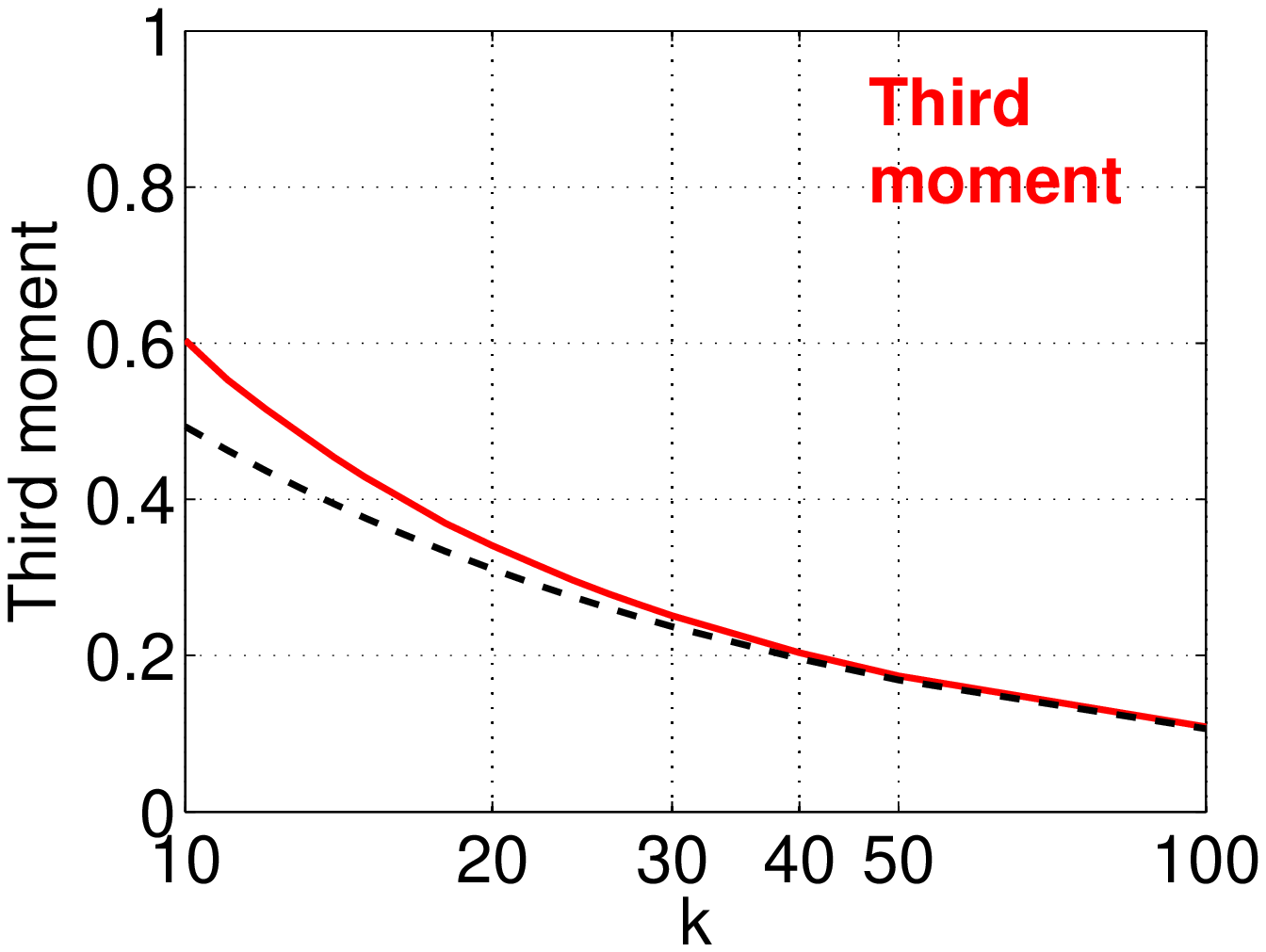}}
\subfigure[{\scriptsize $\left(\text{E}(\hat{d}_{MLE,c}-\text{E}(\hat{d}_{MLE,c}))^4/d^4\right)^{1/4}$}]{\includegraphics[width = 2.25in]{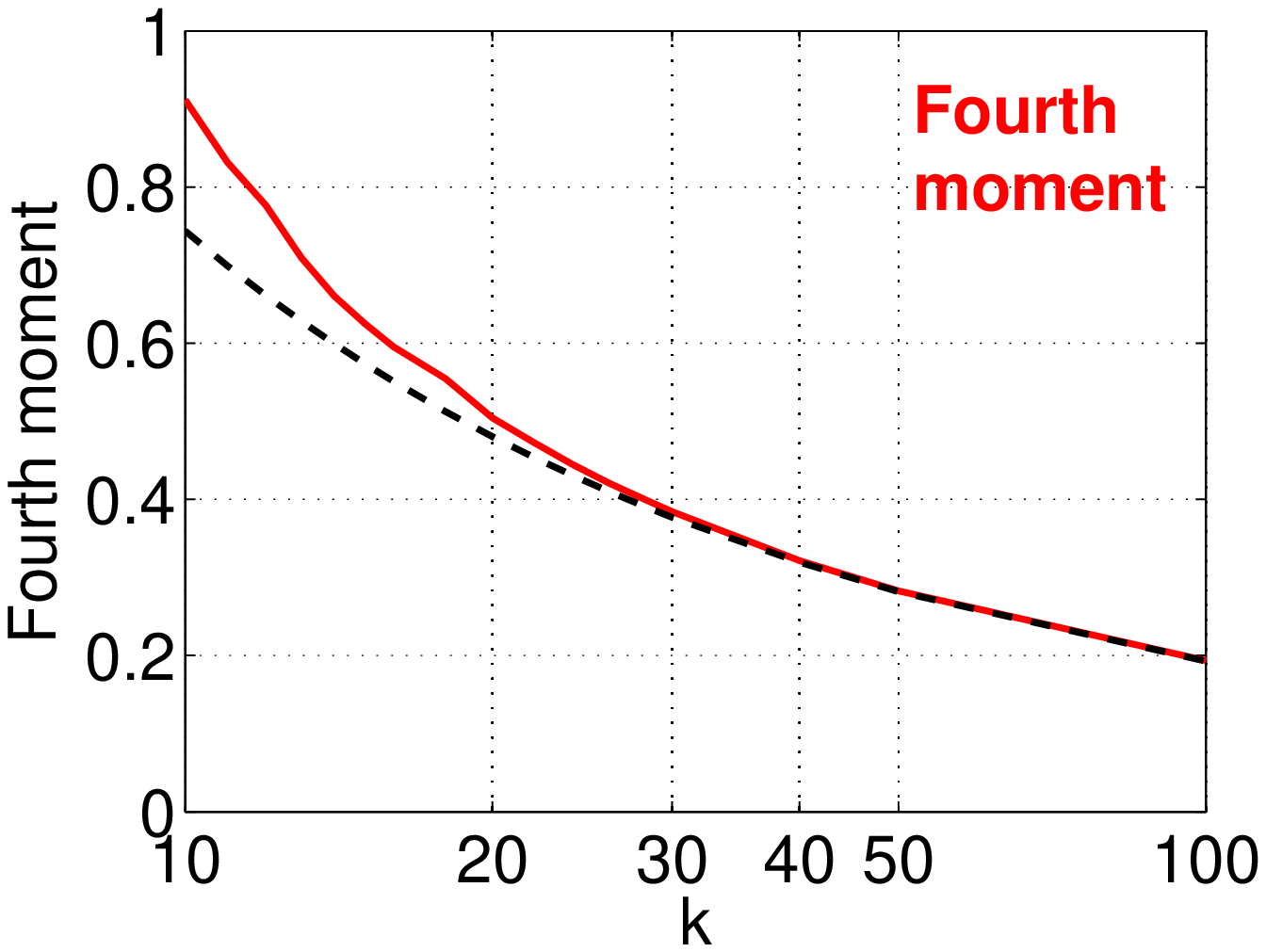}}}
\end{center}\vspace{-0.45in}
\caption{One pair of words are selected from an  MSN term-by-document
  matrix with $D=2^{16}$ Web pages. We conduct Cauchy random
  projections and estimate the $l_1$ distance between one pair of words using the maximum
  likelihood estimator $\hat{d}_{MLE}$ and the bias-corrected version
  $\hat{d}_{MLE,c}$. Panel (a)
  plots the biases of $\hat{d}_{MLE}$ and $\hat{d}_{MLE,c}$, indicating that
  the bias correction is effective. Panels (b), (c), and
  (d) plot the variance, third moment, and fourth moment of
  $\hat{d}_{MLE,c}$, respectively. The dashed curves are the theoretical
  asymptotic moments. When $k\geq 20$,
  the theoretical asymptotic formulas for moments are accurate.}\label{fig_bias_var}\vspace{-0.1in}
\end{figure}

%\vspace{-0.1in}
The maximum likelihood estimators are tested on MSN Web crawl
data, a term-by-document matrix with
$D=2^{16}$ Web pages. We conduct Cauchy random 
projections and estimate the $l_1$ distances
between words.  In this experiment, we compare the empirical and
(asymptotic) theoretical moments, using one pair of words. Figure \ref{fig_bias_var} illustrates that the bias correction is
effective and these (asymptotic) formulas for the first four moments
of $\hat{d}_{MLE,c}$ in Lemma \ref{lem_mle_asymp} are accurate, especially when $k\geq 20$.\vspace{-0.25in}

\subsection{Approximation Distributions}

Theoretical analysis on the exact distribution of a maximum likelihood
estimator is difficult.\footnote{In fact, conditional on the observations $x_1$,
  $x_2$, ..., $x_k$, the distribution of $\hat{d}_{MLE}$ can be exactly
  characterized \citep{Article::Fisher_34}.  \cite{Article:Lawless_72}
  studied the conditional confidence interval of the MLE. Later, 
   \cite{Article:Hinkley_78} proposed the normal approximation to the exact
conditional confidence interval and showed that it was superior to the
unconditional normality approximation. Unfortunately, we can not take advantage of the conditional
analysis because our goal is to determine the sample size $k$ before
seeing any samples. } In statistics, the standard
approach is to assume normality, which, however, is quite
inaccurate. The so-called {\em Edgeworth expansion}\footnote{The so-called {\em Saddlepoint approximation} in general improves
Edgeworth expansions \citep{Book:Jensen_95}, often very
considerably. Unfortunately, we can not apply the Saddlepoint
approximation in our case (at least not directly), because the
Saddlepoint approximation needs a bounded moment generating
function.} improves the
normal approximation by matching higher moments
\citep{Book:Feller_II,Article:Bhattacharya_78, Book:Severini_00}. For
example, if we approximate the distribution of $\hat{d}_{MLE,c}$ using
an Edgeworth expansion by matching the first four moments of
$\hat{d}_{MLE,c}$ derived in Lemma \ref{lem_mle_asymp}, then the errors
 will be on the order of $O\left(k^{-3/2}\right)$. However, Edgeworth
 expansions have some well-known drawbacks. The resultant
 expressions are quite sophisticated. They are not accurate at
 the tails. It is possible that the approximate probability has values
 below zero. Also, Edgeworth expansions consider the support is
 $(-\infty, \infty)$, while  $\hat{d}_{MLE,c}$ is 
 non-negative.

We propose approximating the distributions of
$\hat{d}_{MLE,c}$ directly using some well-studied common
distributions. We will first consider a gamma distribution with the
same first two (asymptotic) moments of $\hat{d}_{MLE,c}$. That is, the
gamma distribution will be asymptotically equivalent to the normal
approximation. While a normal has zero third
central moment, a gamma has nonzero third central moment. This, to an
extent, speeds up the rate of convergence. Another important reason
why a gamma is more accurate is because it has the same support as
$\hat{d}_{MLE,c}$, i.e., $[0,\infty)$. 

We will furthermore consider a {\em   generalized gamma} distribution,
which allows us to match the first 
three (asymptotic) moments of $\hat{d}_{MLE,c}$.  Interestingly, in
this case, the generalized gamma approximation turns out to be an
inverse Gaussian distribution, which has a closed-form probability density. More
interestingly, this inverse Gaussian distribution also 
matches the fourth central moment of $\hat{d}_{MLE,c}$ in the
$O\left(\frac{1}{k^2}\right)$ term and almost in the
$O\left(\frac{1}{k^3}\right)$ term. By simulations, the inverse
Gaussian approximation is highly accurate. 

Note that, since we are interested in the very small (e.g., $10^{-10}$) tail probability
range, $O\left(k^{-3/2}\right)$ is not too meaningful. For example,
$k^{-3/2} = 10^{-3}$ if $k = 100$. Therefore, we will have to
rely on simulations to assess the accuracy of the approximations. On
the other hand, an upper
bound may hold exactly (verified by simulations) even if it is based
on an approximate distribution. 

As the related work, \cite{Article:Li_SINR06} applied gamma and generalized gamma 
approximations to model the performance measure distribution in some
wireless communication channels using random matrix theory and
produced  accurate results in evaluating the error probabilities. 
 
\subsubsection{The Gamma Approximation}

The gamma approximation is an obvious improvement over the normal
approximation.\footnote{In {\em normal random projections} for
  dimension reduction in $l_2$, the resultant estimator of the squared
  $l_2$
  distance has a chi-squared distribution (e.g., \cite[Lemma
  1.3]{Book:Vempala}), which is a special case of gamma.} 
A gamma distribution, $G(\alpha,\beta)$, has two parameters, $\alpha$
and $\beta$, which can be determined by matching the first two
(asymptotic) moments of $\hat{d}_{MLE,c}$. That is, we assume that $\hat{d}_{MLE,c} \sim G(\alpha, \beta)$, with 
\begin{align}
&\alpha\beta = d, \hspace{0.25in} \alpha\beta^2 = \frac{2d^2}{k} +
\frac{3d^2}{k^2}, \ \ \ 
\Longrightarrow \  \
\alpha = \frac{1}{\frac{2}{k} + \frac{3}{k^2}}, \hspace{0.25in} \beta = \frac{2d}{k} + \frac{3d}{k^2}.
\end{align}

Assuming a gamma distribution, it is easy to obtain the following
Chernoff bounds\footnote{Using the Chernoff inequality
  \citep{Article:Chernoff_52}, we bound the tail probability by 
$\mathbf{Pr}\left(Q>z\right) = \mathbf{Pr}\left(e^{Qt}>e^{zt}\right)
\leq \text{E}\left(e^{Qt}\right)e^{-zt}$; and we then choose $t$ that minimizes
the upper bound.}: 
\begin{align}\label{eqn_gamma_right}
&\mathbf{Pr}\left(\hat{d}_{MLE,c} \geq  (1+\epsilon)
  d\right)  \overset{\sim}{\leq} \exp\left(-\alpha\left(\epsilon -
    \log(1+\epsilon)\right)\right), \hspace{0.2in} \epsilon \geq 0 \\
&\mathbf{Pr}\left(\hat{d}_{MLE,c} \leq (1-\epsilon)
  d\right)  \overset{\sim}{\leq} \exp\left(-\alpha\left(-\epsilon -
    \log(1-\epsilon)\right)\right), \hspace{0.2in} 0\leq \epsilon <
1\label{eqn_gamma_left}, 
\end{align}
\noindent where we use $\overset{\sim}{\leq}$ to indicate that these
inequalities are based on an approximate distribution. 

\begin{figure}[h]
\begin{center}\mbox{
\subfigure[]{\includegraphics[width = 2.8in]{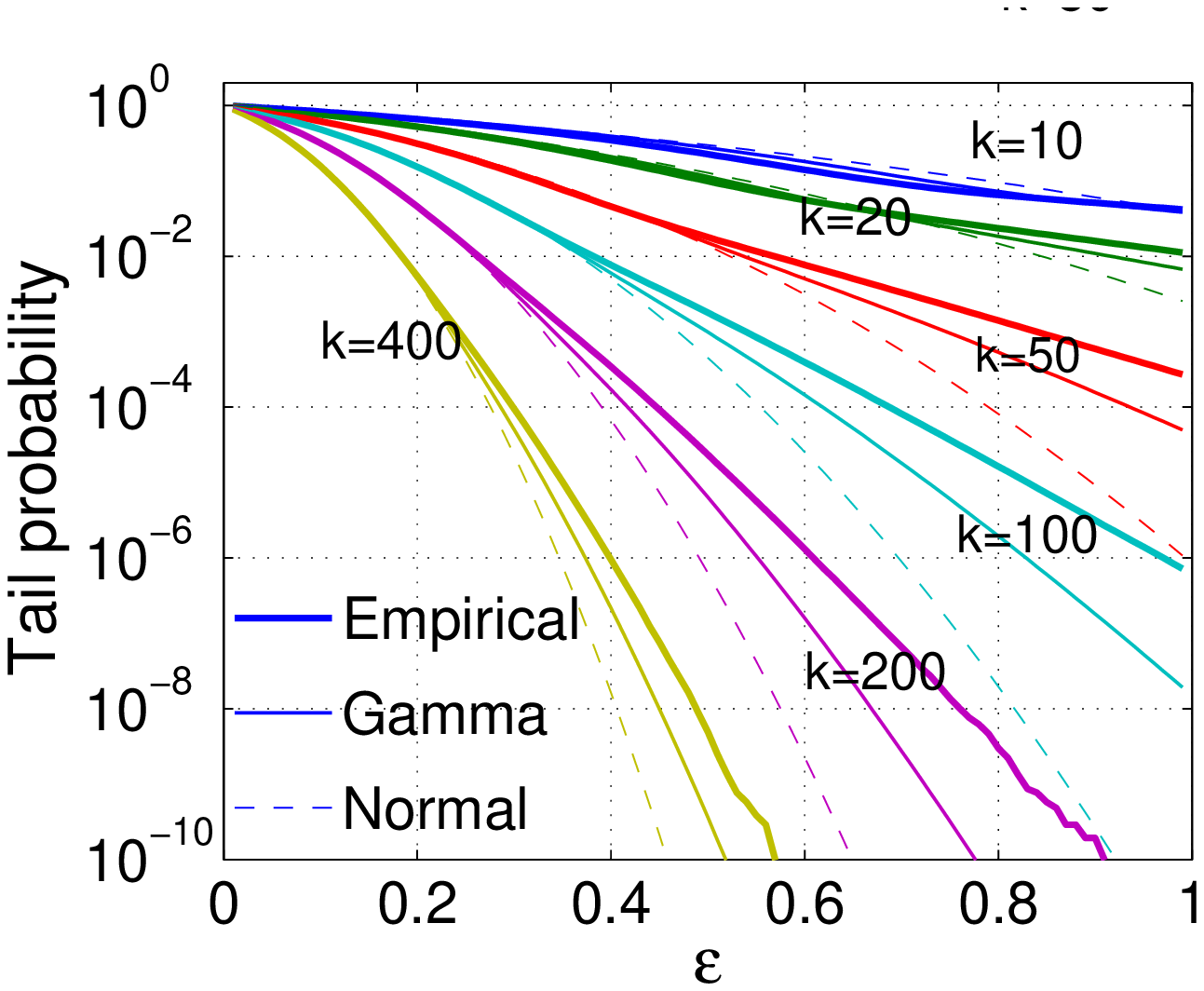}}
\subfigure[]{\includegraphics[width = 2.8in]{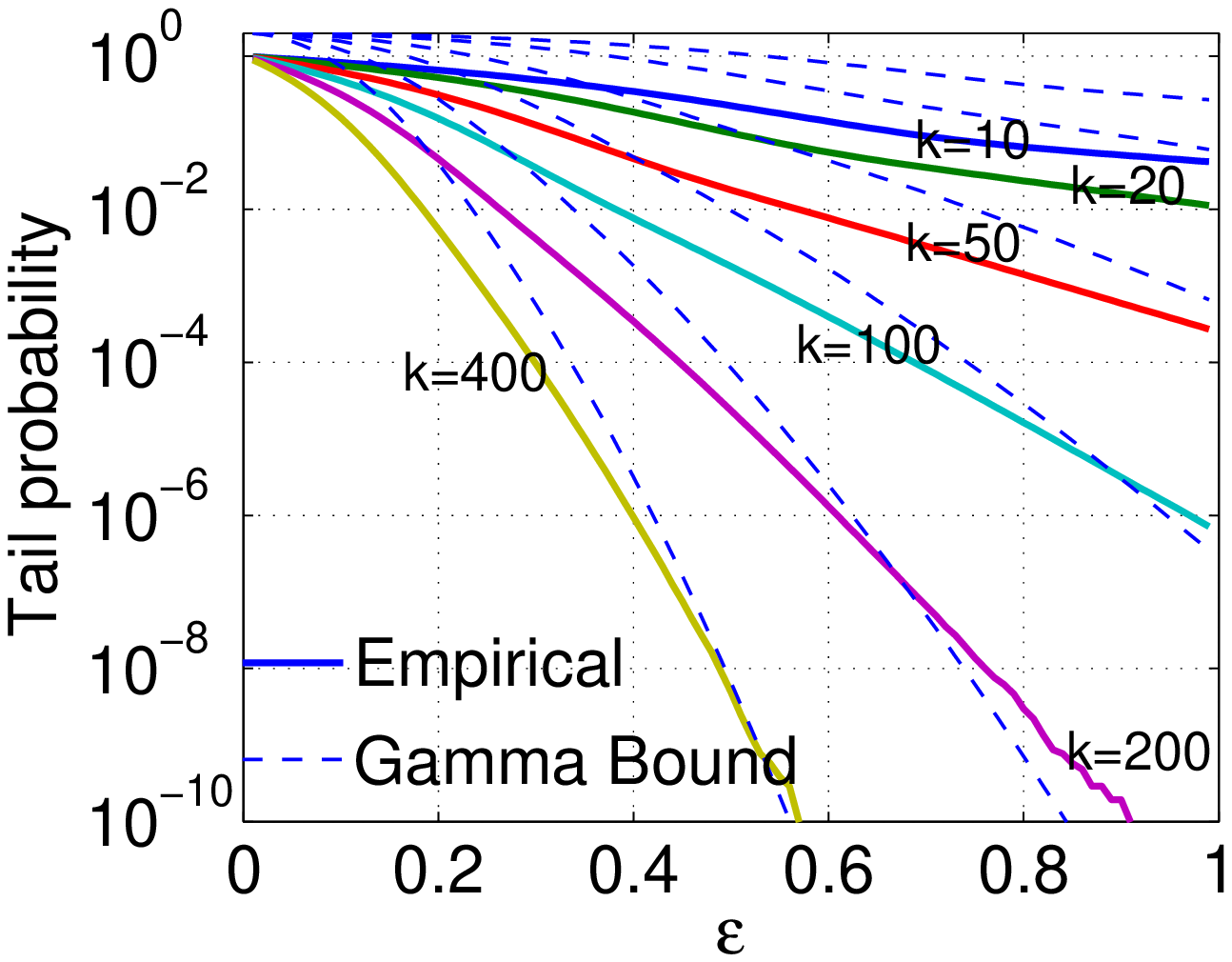}}}
\end{center}\vspace{-0.4in}
\caption{ We consider $k$ = 10, 20, 50, 100, 200, and 400. For each $k$, we
  simulate standard Cauchy samples, from which we
  estimate the Cauchy parameter by the MLE $\hat{d}_{MLE,c}$ and compute the tail
probabilities. Panel (a) compares the empirical tail probabilities
(thick solid) with
the gamma tail probabilities (thin solid), indicating that the gamma distribution
is better than the
normal  (dashed) for approximating the distribution of
$\hat{d}_{MLE,c}$.  Panel (b) compares the empirical tail
probabilities with the gamma upper bound
(\ref{eqn_gamma_right})+(\ref{eqn_gamma_left}).  }\label{fig_gamma_tail}
\end{figure}

Note that the distribution of $\hat{d}_{MLE}/d$ (and hence $\hat{d}_{MLE,c}/d$) is only a function of
$k$ as shown in \citep{Article:Antle_69,Article:Haas_70}. Therefore, we
can evaluate the accuracy of the gamma approximation by simulations
with $d = 1$, as presented in Figure \ref{fig_gamma_tail}.

Figure \ref{fig_gamma_tail}(a) shows that both the gamma and
normal approximations are fairly accurate when the tail probability $\geq
10^{-2}\sim 10^{-3}$; and the gamma approximation is  obviously
better. 

Figure \ref{fig_gamma_tail}(b) compares the empirical tail probabilities with the 
gamma Chernoff upper bound
(\ref{eqn_gamma_right})+(\ref{eqn_gamma_left}), indicating that these bounds are reliable, when the tail probability $\geq
10^{-5}\sim 10^{-6}$.

\subsubsection{The Inverse Gaussian  (Generalized Gamma) Approximation}

The distribution of $\hat{d}_{MLE,c}$ can be well
approximated by an inverse Gaussian distribution, which is a special
case of the three-parameter generalized gamma distribution
 \citep{Article:Hougaard_86,Article:Gerber}, denoted by $GG(\alpha, \beta,
\eta)$. Note that the usual gamma distribution is a special case
with $\eta = 1$. 

If $z \sim GG(\alpha, \beta, \eta)$, then the first
three moments are 
\begin{align}
\text{E}(z) = \alpha\beta, \hspace{0.2in} \text{Var}(z) =
\alpha\beta^2, \hspace{0.2in} \text{E}\left(z - \text{E}(z)\right)^3 =
\alpha\beta^3(1+\eta). 
\end{align}

We can approximate the distribution of $\hat{d}_{MLE,c}$ by matching the
first three moments, i.e., 
\begin{align}
\alpha\beta = d, \hspace{0.2in} \alpha\beta^2 = \frac{2d^2}{k} +
\frac{3d^2}{k^2}, \hspace{0.2in} \alpha\beta^3(1+\eta) =
\frac{12d^3}{k^2}, 
\end{align}
\noindent from which we obtain
\begin{align}
\alpha = \frac{1}{\frac{2}{k} + \frac{3}{k^2}}, \hspace{0.2in} \beta
= \frac{2d}{k} + \frac{3d}{k^2}, \hspace{0.2in} \eta = 2 +
O\left(\frac{1}{k}\right). \label{eqn_ig_parameters}
\end{align}
Taking only the leading term for $\eta$, the generalized gamma
approximation of $\hat{d}_{MLE,c}$ would be 
\begin{align}
GG\left(\frac{1}{\frac{2}{k} + \frac{3}{k^2}}, \frac{2d}{k} +
  \frac{3d}{k^2}, 2\right). \label{eqn_ig}
\end{align}

In general, a generalized gamma distribution does not have a closed-form
density function although it always has a closed-from moment generating
function.  In our case, (\ref{eqn_ig}) is actually an
inverse Gaussian distribution, which has a closed-form density
function. Assuming $\hat{d}_{MLE,c} \sim IG(\alpha, \beta)$,
with parameters $\alpha$ and
$\beta$ defined in (\ref{eqn_ig_parameters}), the moment
generating function (MGF), the probability density
function (PDF), and cumulative density function (CDF) would
be \citep[Chapter 2]{Book:Seshadri_93} \citep{Article:Tweedie_57I,Article:Tweedie_57II}\footnote{The inverse Gaussian distribution was first noted as the
  distribution of the first passage time of the Brownian motion with a
  positive drift. It has many interesting properties such as
  infinitely divisible. Two monographs
   \citep{Book:Chhikara_89,Book:Seshadri_93} are devoted entirely to the
  inverse Gaussian distributions. For a quick reference, one can check
{\it http://mathworld.wolfram.com/InverseGaussianDistribution.html}.}
\begin{align}
&\text{E}\left(\exp(\hat{d}_{MLE,c}t)\right) \overset{\sim}{=}
\exp\left(\alpha\left(1-(1-2\beta t)^{1/2}\right)\right),\\
&\mathbf{Pr}(\hat{d}_{MLE,c} = y)\overset{\sim}{=} \frac{\alpha \sqrt{\beta}}{\sqrt{2\pi}}
y^{-\frac{3}{2}} \exp\left(-\frac{\left(y/\beta -
      \alpha\right)^2}{2y/\beta}\right) = \sqrt{\frac{\alpha d}{2\pi}}y^{-\frac{3}{2}} \exp\left(-\frac{\left(y-d\right)^2}{2y\beta}\right),\\ \notag
&\mathbf{Pr}\left(\hat{d}_{MLE,c} \leq y\right) \overset{\sim}{=}
\Phi\left(\sqrt{\frac{\alpha^2\beta}{y}}\left(\frac{y}{\alpha\beta} -1
    \right)\right) + e^{2\alpha}
  \Phi\left(-\sqrt{\frac{\alpha^2\beta}{y}}\left(\frac{y}{\alpha\beta}
      +1 
    \right)\right)\\
&\hspace{1.1in}= 
\Phi\left(\sqrt{\frac{\alpha d}{y}}\left(\frac{y}{d} -1
    \right)\right) + e^{2\alpha}
  \Phi\left(-\sqrt{\frac{\alpha d}{y}}\left(\frac{y}{d}
      +1 
    \right)\right), 
\end{align}
\noindent where $\Phi(.)$ is the standard normal CDF, i.e., $\Phi(z) =
\int_{-\infty}^z \frac{1}{\sqrt{2\pi}}e^{-\frac{t^2}{2}}dt$. Here we
use $\overset{\sim}{=}$ to indicate that these equalities are based on
an approximate distribution.

Assuming $\hat{d}_{MLE,c} \sim
IG(\alpha,\beta)$, then the fourth central moment should be 
\begin{align}\notag
\text{E}\left(\hat{d}_{MLE,c} - \text{E}\left(\hat{d}_{MLE,c}\right)\right)^4 &\overset{\sim}{=}
15\alpha\beta^4+ 3\left(\alpha\beta^2\right)^2 \\\notag
&=15d\left(\frac{2d}{k}+\frac{3d}{k^2}\right)^3 +
3\left(\frac{2d^2}{k}+\frac{3d^2}{k^2}\right)^2 \\
&=\frac{12d^4}{k^2} + \frac{156d^4}{k^3} +
O\left(\frac{1}{k^4}\right). 
\end{align}

Lemma \ref{lem_mle_asymp} has shown the true asymptotic fourth central
moment: 
\begin{align}
\text{E}\left(\hat{d}_{MLE,c} -
  \text{E}\left(\hat{d}_{MLE,c}\right)\right)^4 =\frac{12d^4}{k^2} + \frac{186d^4}{k^3} +
O\left(\frac{1}{k^4}\right).
\end{align}
\noindent That is, the inverse Gaussian approximation matches not only the
leading term, $\frac{12d^4}{k^2}$, but also almost the higher
order term, $\frac{186d^4}{k^3}$, of the true asymptotic fourth moment of
 $\hat{d}_{MLE,c}$.

Assuming $\hat{d}_{MLE,c} \sim IG(\alpha,\beta)$, the tail probability
of $\hat{d}_{MLE,c}$ can be expressed  as 
\begin{align}
&\mathbf{Pr}\left(\hat{d}_{MLE,c} \geq (1+\epsilon)d\right) \overset{\sim}{=}
\Phi\left(-\epsilon \sqrt{\frac{\alpha}{1+\epsilon}}\right) -
e^{2\alpha} \Phi\left(-(2+\epsilon)\sqrt{\frac{\alpha}{1+\epsilon}}\right),
\hspace{0.1in} \epsilon \geq 0 \\
&\mathbf{Pr}\left(\hat{d}_{MLE,c} \leq (1-\epsilon)d\right)  \overset{\sim}{=} \Phi\left(-\epsilon \sqrt{\frac{\alpha}{1-\epsilon}}\right) +
e^{2\alpha} \Phi\left(-(2-\epsilon)\sqrt{\frac{\alpha}{1-\epsilon}}\right),
\hspace{0.1in}   0\leq \epsilon < 1. 
\end{align}

Assuming  $\hat{d}_{MLE,c} \sim IG(\alpha,\beta)$, it is easy to show
the following  Chernoff bounds: 
\begin{align}\label{eqn_ig_left}
&\mathbf{Pr}\left(\hat{d}_{MLE,c} \geq (1+\epsilon)d\right) \overset{\sim}{\leq}
\exp\left(-\frac{\alpha \epsilon^2}{2(1+\epsilon)}\right),  \hspace{0.2in} \epsilon \geq 0 \\
&\mathbf{Pr}\left(\hat{d}_{MLE,c} \leq (1-\epsilon)d\right) \overset{\sim}{\leq}
\exp\left(-\frac{\alpha \epsilon^2}{2(1-\epsilon)}\right),
\hspace{0.2in}   0\leq \epsilon < 1. \label{eqn_ig_right}
\end{align}

To see (\ref{eqn_ig_left}). Assume $z \sim IG(\alpha,\beta)$. Then,
using the Chernoff inequality: 
\begin{align}\notag
\mathbf{Pr}\left(z \geq (1+\epsilon)d\right) \leq&
\text{E}\left(zt\right)\exp(-(1+\epsilon)dt)\\\notag
=&\exp\left(\alpha\left(1-(1-2\beta t)^{1/2}\right)-(1+\epsilon)dt\right),
\end{align}
whose minimum is $\exp\left(-\frac{\alpha
    \epsilon^2}{2(1+\epsilon)}\right)$, attained at $t =
\left(1-\frac{1}{(1+\epsilon)^2}\right)\frac{1}{2\beta}$. We can
similarly show (\ref{eqn_ig_right}). \\

Combining (\ref{eqn_ig_left}) and (\ref{eqn_ig_right}) yields a
symmetric bound 
\begin{align}
&\mathbf{Pr}\left(|\hat{d}_{MLE,c} - d| \geq \epsilon d\right) \overset{\sim}{\leq}
2\exp\left(-\frac{\epsilon^2/(1+\epsilon)}{2 \left(\frac{2}{k} + \frac{3}{k^2}\right)}\right),
\hspace{0.15in} 0\leq \epsilon \leq 1
\end{align}

\begin{figure}[h]
\begin{center}\mbox{
\subfigure[]{\includegraphics[width = 2.8in]{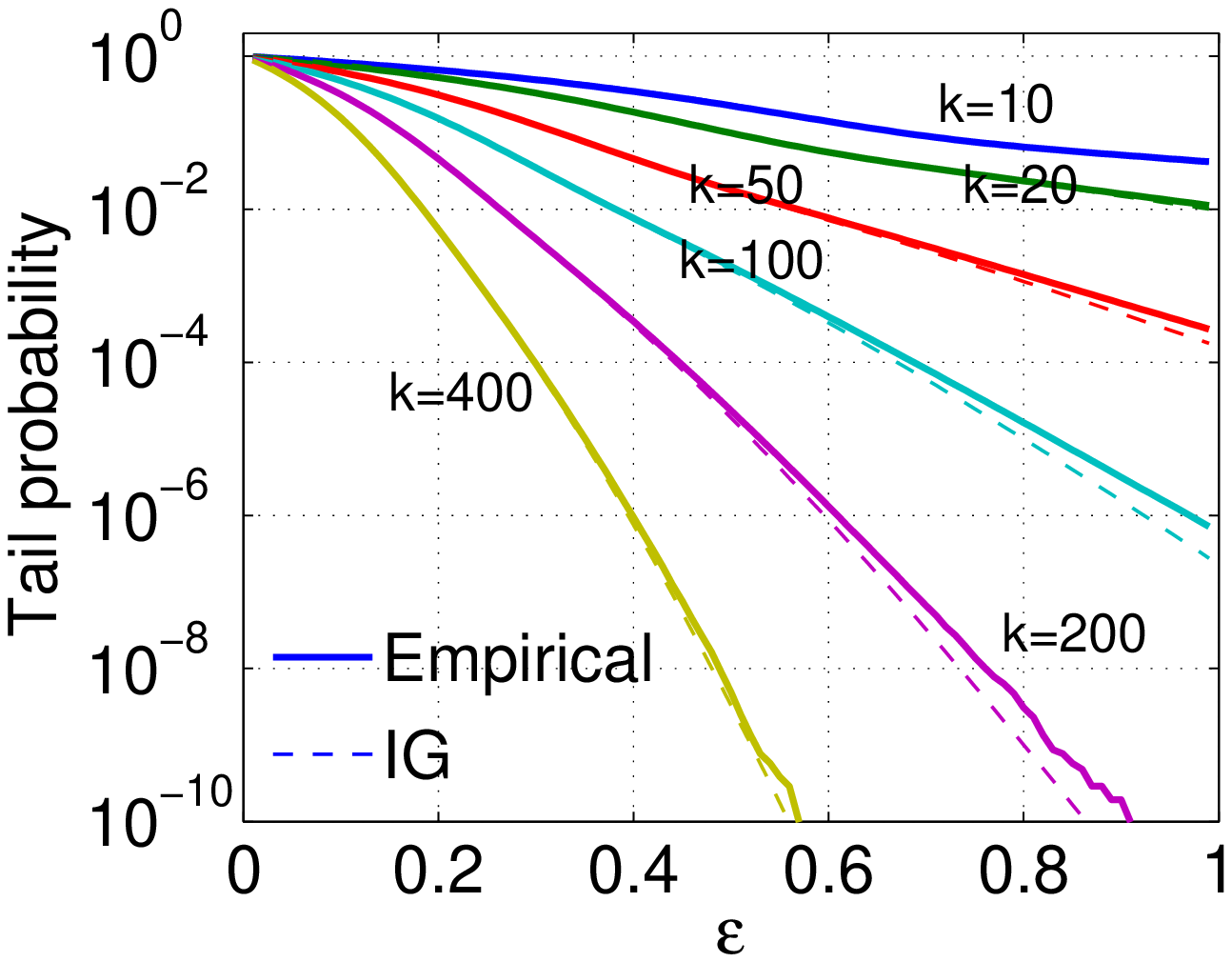}}
\subfigure[]{\includegraphics[width = 2.8in]{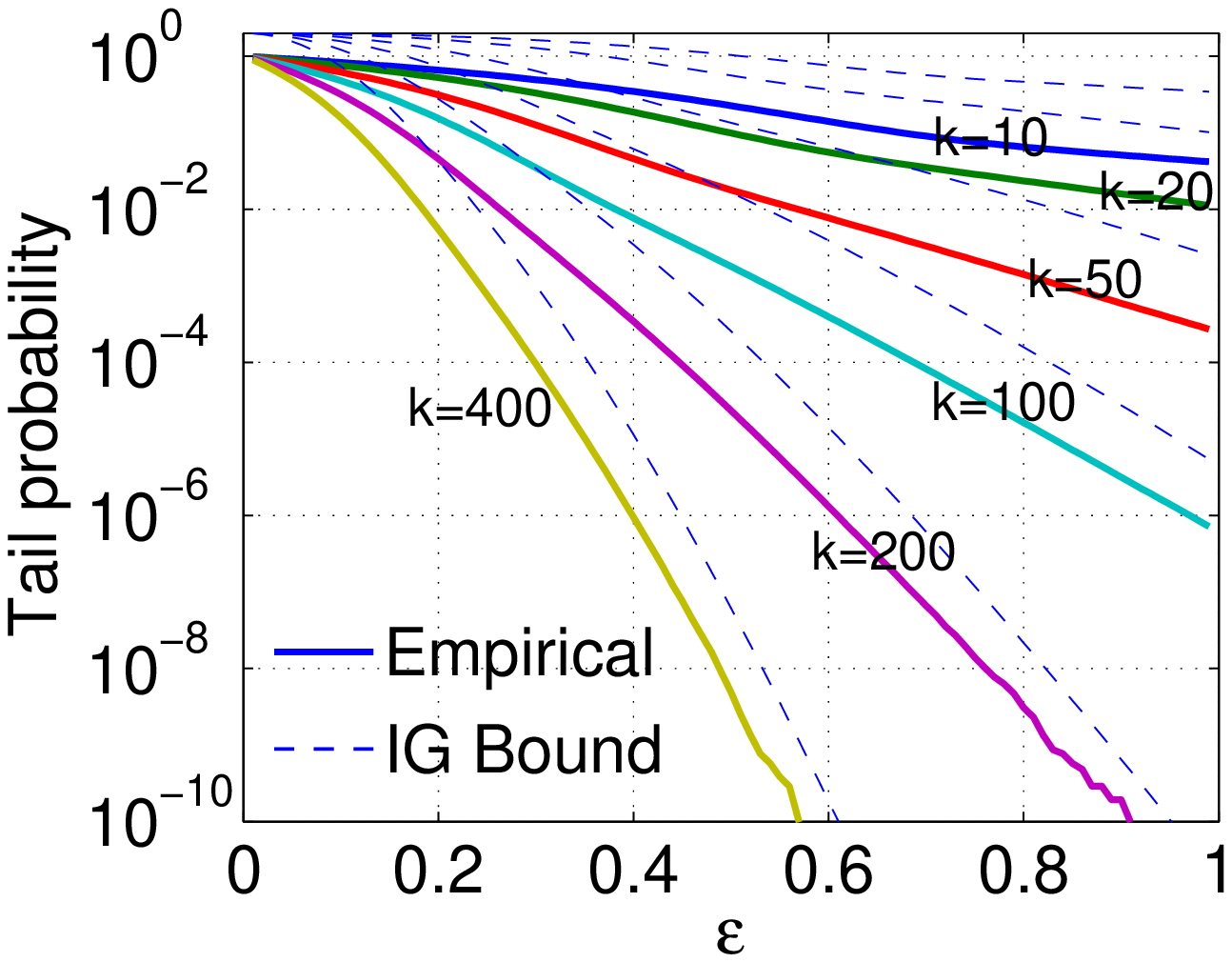}}}
\end{center}\vspace{-0.4in}
\caption{We compare the inverse Gaussian approximation
  with the same simulations as presented in Figure
  \ref{fig_gamma_tail}. Panel (a) compares the empirical tail
  probabilities with the inverse Gaussian tail probabilities,
  indicating that the approximation is highly accurate. 
  Panel (b) compares the empirical tail probabilities with the inverse
  Gaussian upper bound (\ref{eqn_ig_left})+(\ref{eqn_ig_right}). The upper bounds are all
above the corresponding empirical curves, indicating that our proposed bounds are
reliable at least in our simulation range.  }\label{fig_ig_tail}
\end{figure}

Figure \ref{fig_ig_tail} compares the inverse Gaussian approximation with the same
simulations as presented in Figure \ref{fig_gamma_tail}, indicating 
that the inverse Gaussian approximation is highly
accurate. When the tail probability $\geq 10^{-4} \sim 10^{-6}$, we can treat the
inverse Gaussian as the exact distribution of $\hat{d}_{MLE,c}$.  The Chernoff upper bounds for the inverse Gaussian
are always reliable in our simulation range (the tail probability
$\geq 10^{-10}$). 

\section{Conclusion}\label{sec_conclusion}

It is well-known that the $l_1$ distance is far more robust than the
$l_2$ distance against ``outliers.'' There are
numerous  success stories of using the $l_1$ distance, e.g., 
  Lasso \citep{Article:Tibshirani_96}, LARS \citep{Article:Efron_LARS04}, 1-norm
  SVM \citep{Proc:Zhu_NIPS03}, and Laplacian radial basis kernel
  \citep{Article:Chapelle_99,Proc:Ferecatu_MIR04}. 

Dimension reduction in the $l_1$ norm, however, has been proved
{\em impossible} if we use {\em linear random projections} and {\em
  linear estimators}. In this study, we propose three types of nonlinear
estimators for {\em Cauchy random projections}: the bias-corrected
sample median estimator, the bias-corrected geometric mean estimator,
and the bias-corrected maximum likelihood estimator. Our theoretical
analysis has shown that these nonlinear estimators can accurately
recover the original $l_1$ distance, even though none of them can be a
metric. 

The bias-corrected sample median estimator and the bias-corrected
geometric mean estimator are asymptotically equivalent but the latter
is more accurate at small sample size. We have derived explicit tail
bounds for the bias-corrected geometric mean estimator and have expressed
the tail bounds in exponential forms. Using these tail bounds, we have
established an analog of the 
Johnson-Lindenstrauss  (JL) lemma for dimension reduction in $l_1$, which is weaker than the classical JL lemma for dimension reduction in
$l_2$.

We conduct theoretic analysis  on the bias-corrected maximum
likelihood estimator (MLE), which is ``asymptotically optimum.'' Both
the sample median estimator and the geometric mean estimator are about
$80\%$ efficient as the MLE. We propose
approximating its distribution by an inverse Gaussian, which has the
same support and matches the leading terms of the first four moments of
the proposed estimator. Approximate tail bounds have been provide based
on the inverse Gaussian approximation. Verified by simulations, these
approximate tail bounds hold at least in the $\geq 
10^{-10}$ tail probability range. 

Although these nonlinear estimators are not metrics, they are still
useful for certain applications in (e.g.,) data stream computation,
information retrieval, learning and data mining, whenever the goal is
to compute the $l_1$ distances efficiently using a small storage space.

The geometric mean estimator is a non-convex
norm (i.e., the $l_p$ norm as $p\rightarrow 0$); and therefore it does
contain some information about the geometry.  It may be still possible
to develop certain efficient algorithms using the geometric mean estimator by
avoiding the non-convexity.  We leave this for future
work. \\

\section*{Acknowledgment}

We are grateful to Piotr Indyk and Assaf Naor for the very constructive
comments on various versions of this manuscript. We thank Dimitris
Achlioptas, 
Christopher Burges, Moses Charikar, Jerome Friedman, Tze L. Lai, Art 
B. Owen, John Platt,  Joseph Romano, Tim
Roughgarden, Yiyuan She,  and  Guenther Walther
for helpful conversations or suggesting relevant references. We also thank Silvia Ferrari 
and Gauss Cordeiro for clarifying some parts of their papers. 

Trevor Hastie was partially supported by grant DMS-0505676 from the National
Science Foundation, and grant 2R01 CA 72028-07 from the National
Institutes of
Health.

{\small

%\bibliography{../bib/IEEEabrv,../bib/mybibfile}
}

\appendix

\section{Proof of Lemma \ref{lem_me}}\label{app_proof_lem_me}

Assume $x \sim C(0,d)$. The probability density function (PDF) and the
cumulative density function (CDF) of $|x|$ would be 
\begin{align}
&\mathbf{Pr}(|x|=z) = \frac{2d}{\pi}\frac{1}{z^2+d^2}, \hspace{0.2in}
z\geq0 \\
&\mathbf{Pr}(|x|\leq z) = \frac{2}{\pi}\tan^{-1}\frac{z}{d}, \hspace{0.2in}
z\geq0
\end{align}

The asymptotic normality of $\hat{d}_{me}$ follows from the asymptotic
results on sample quantiles \citep[Theorem
5.10]{Book:Shao}.
\begin{align}
\sqrt{k}\left(\hat{d}_{me}-d\right) \overset{D}{\Longrightarrow}
N\left(0,
  \frac{1}{2}\left(1-\frac{1}{2}\right)/\left(\left.\mathbf{Pr}(|x|=z)\right|_{z = d}\right)^2\right) = N\left(0,\frac{\pi^2}{4}d^2\right)
\end{align}

The probability density of $\hat{d}_{me}$ can be derived from
the probability density of order statistics \citep[Example
2.9]{Book:Shao}. For simplicity, we only consider $k = 2m+1$, $m = 1,
2, ..., $
\begin{align}\notag
\mathbf{Pr}(\hat{d}_{me}=z) &=
\frac{(2m+1)!}{(m!)^2}\left(\mathbf{Pr}(|x|\leq
  z)\right)^m\left(1-\mathbf{Pr}(|x|\leq z)\right)^m \mathbf{Pr}(|x|=
z) \\
&=\frac{(2m+1)!}{(m!)^2}\left(\frac{2}{\pi}\tan^{-1}\frac{z}{d}\right)^m\left(1-\frac{2}{\pi}\tan^{-1}\frac{z}{d}\right)^m\frac{2d}{\pi}\frac{1}{z^2+d^2}.
\end{align}

The $r^{th}$ moment of $\hat{d}_{me}$ would be 
\begin{align}\notag
\text{E}\left(\hat{d}_{me}\right)^r &= \int_0^\infty z^r
\frac{(2m+1)!}{(m!)^2}\left(\frac{2}{\pi}\tan^{-1}\frac{z}{d}\right)^m\left(1-\frac{2}{\pi}\tan^{-1}\frac{z}{d}\right)^m\frac{2d}{\pi}\frac{1}{z^2+d^2}dz
\\
&= d^r\int_0^1\frac{(2m+1)!}{(m!)^2}\tan^r\left(\frac{\pi}{2}t\right)
\left(t-t^2\right)^m dt,
\end{align}
\noindent by substituting $t = \frac{2}{\pi}
\tan^{-1}\frac{z}{d}$.

When $t\rightarrow 1-0$, $\tan\left(\frac{\pi}{2}t\right) \rightarrow
\infty$, but $t-t^2 = t(1-t) \rightarrow 0$. Around $t =1-0$,  
 $\tan\left(\frac{\pi}{2}t\right) =
 \frac{1}{\tan\left(\frac{\pi}{2}(1-t)\right)} =
 \frac{2}{\pi}\frac{1}{1-t}+...$, by the Taylor expansion. Therefore, in
 order for $\text{E}\left(\hat{d}_{me}\right)^r <\infty$, we must have
 $m \geq r$.

We complete the proof of Lemma \ref{lem_me}. 

\section{Proof of  Lemma \ref{lem_d_log}}\label{app_proof_lem_d_log} 

Assume $x \sim C(0,d)$. The first moment of $\log(|x|)$ would be
\begin{align}\notag
\text{E}\left(\log(|x|)\right) &= \frac{2d}{\pi}\int_0^\infty
\frac{\log(y)}{y^2+d^2}dy \\\notag
&=\frac{1}{\pi}\int_0^\infty\frac{\log(d)y^{-1/2}}{y+1} +
\frac{1/2\log(y)y^{-1/2}}{y+1}dy\\
&= \log(d), 
\end{align}
\noindent with the help of the integral tables \cite[3.221.1,
4.251.1]{Book:Gradshteyn_94}. 

Thus, given i.i.d. samples $x_j \sim C(0,d)$, $j = 1, 2, ..., k$,
a nonlinear estimator of $d$ would be 
\begin{align}
\hat{d}_{log} = \exp\left(\frac{1}{k}\sum_{j=1}^k\log(|x_j|)\right). 
\end{align}

We can derive another nonlinear estimator from 
$\text{E}\left(|x|^\lambda\right)$, $|\lambda| <1$. Using the integral
tables \cite[3.221.1]{Book:Gradshteyn_94}, we obtain
\begin{align}\notag
\text{E}\left(|x|^\lambda\right) &= \frac{2d}{\pi}\int_0^\infty
\frac{y^\lambda}{y^2+d^2}dy\\ \notag
&=\frac{d^\lambda}{\pi}\int_0^\infty\frac{y^{\frac{\lambda-1}{2}}}{y+1}dy
\\
&=\frac{d^\lambda}{\cos(\lambda\pi/2)}, 
\end{align}
\noindent from which a  nonlinear estimator follows immediately
\begin{align}
\hat{d}_\lambda = \left(\frac{1}{k}\sum_{j=1}^k|x_j|^\lambda
  \cos(\lambda\pi/2)\right)^{1/\lambda}, \hspace{0.2in} |\lambda| <1
\end{align}

Both nonlinear estimators $\hat{d}_{log}$ and $\hat{d}_\lambda$ are
biased. The leading terms of their variances can be obtained by the
{\em Delta Method} \citep[Corollary 1.1]{Book:Shao}.

With the help of \cite[4.261.10]{Book:Gradshteyn_94}, we obtain 
\begin{align}
\text{E}\left(\log^2(|x|)\right) = \log^2(d) + \frac{\pi^2}{4},
\hspace{0.2in} \text{i.e., } \ \ \text{Var}\left(\log^2(|x|)\right) =  \frac{\pi^2}{4}.
\end{align}
\noindent Thus, 
\begin{align}
\text{E}\left(\frac{1}{k}\sum_{j=1}^k\log(|x_j|)\right) = \log d,
\hspace{0.5in} \text{Var}\left(\frac{1}{k}\sum_{j=1}^k\log(|x_j|)\right) = \frac{1}{k}\frac{\pi^2}{4}.
\end{align}

By the {\em Delta Method}, the asymptotic variance of
$\hat{d}_{log}$ should be 
\begin{align}
\text{Var}\left(\hat{d}_{log}\right) =
\frac{1}{k}\frac{\pi^2}{4}\exp^2\left(\log(d)\right) +
O\left(\frac{1}{k^2}\right) = \frac{\pi^2d^2}{4k} +
O\left(\frac{1}{k^2}\right). 
\end{align}

Similarly, the asymptotic variance of $\hat{d}_\lambda$ is 
\begin{align}
\text{Var}\left(\hat{d}_{\lambda}\right) = \frac{d^2}{k}
\frac{\sin^2(\lambda \pi/2)}{\lambda^2 \cos(\lambda\pi)} +
O\left(\frac{1}{k^2}\right), \hspace{0.2in} |\lambda| <1/2
\end{align}

$\text{Var}\left(\hat{d}_{\lambda}\right)\rightarrow \infty$
as $|\lambda|\rightarrow \frac{1}{2}$. $\text{Var}\left(\hat{d}_{\lambda}\right)$
converges to $\text{Var}\left(\hat{d}_{log}\right)$ as $\lambda
\rightarrow 0$, because 
\begin{align}
\underset{\lambda\rightarrow 0}\lim\frac{\sin^2(\lambda
  \pi/2)}{\lambda^2 \cos(\lambda\pi)} = \frac{\pi^2}{4}.
\end{align}

This completes the proof of Lemma \ref{lem_d_log}.

\section{Proof of Lemma \ref{lem_d_gm}}\label{app_proof_lem_d_gm} 

Assume that $x_1$, $x_2$, ..., $x_k$, are i.i.d. $C(0,d)$. 
The estimator, $\hat{d}_{gm,c}$, expressed as
\begin{align}
\hat{d}_{gm,c} = \cos^k\left(\frac{\pi}{2k}\right)\prod_{j=1}^k|x_j|^{1/k},
\end{align}
is unbiased, because, from Lemma \ref{lem_d_log}, 
\begin{align}\notag
\text{E}\left(\hat{d}_{gm,c}\right) &=
  \cos^k\left(\frac{\pi}{2k}\right)\prod_{j=1}^k\text{E}\left(|x_j|^{1/k}\right) \\\notag
&=\cos^k\left(\frac{\pi}{2k}\right)\prod_{j=1}^k\left(\frac{d^{1/k}}{\cos\left(\frac{\pi}{2k}\right)}\right)\\
&=d.
\end{align}

The variance  is 
\begin{align}\notag
\text{Var}\left(\hat{d}_{gm,c}\right) &=
\cos^{2k}\left(\frac{\pi}{2k}\right)\prod_{j=1}^k\text{E}\left(|x_j|^{2/k}\right)
  -d^2\\
&=
d^2
\left(\frac{\cos^{2k}\left(\frac{\pi}{2k}\right)}{\cos^k\left(\frac{\pi}{k}\right)}-1
\right)\\ 
&=\frac{\pi^2}{4}\frac{d^2}{k}  + \frac{\pi^4}{32}\frac{d^2}{k^2}+ O\left(\frac{1}{k^3}\right),
\end{align}
\noindent because
\begin{align}\notag
\frac{\cos^{2k}\left(\frac{\pi}{2k}\right)}{\cos^k\left(\frac{\pi}{k}\right)}
&=
\left(\frac{1}{2}+\frac{1}{2}\left(\frac{1}{\cos(\pi/k)}\right)\right)^k
\\ \notag
&=\left(1+\frac{1}{4}\frac{\pi^2}{k^2} +
  \frac{5}{48}\frac{\pi^4}{k^4}+O\left(\frac{1}{k^6}\right)\right)^k
\\\notag
&=1+k\left(\frac{1}{4}\frac{\pi^2}{k^2}+\frac{5}{48}\frac{\pi^4}{k^4}\right)
+
\frac{k(k-1)}{2}\left(\frac{1}{4}\frac{\pi^2}{k^2}+\frac{5}{48}\frac{\pi^4}{k^4}\right)^2+
... \\
&=1+\frac{\pi^2}{4}\frac{1}{k}+\frac{\pi^4}{32}\frac{1}{k^2} +O\left(\frac{1}{k^3}\right).
\end{align}

Some more algebra can similarly show the third and fourth central moments: 
\begin{align}
&\text{E}\left(\hat{d}_{gm,c} -
  \text{E}\left(\hat{d}_{gm,c}\right)\right)^3 =
\frac{3\pi^4}{16}\frac{d^3}{k^2} + O\left(\frac{1}{k^3}\right)\\
&\text{E}\left(\hat{d}_{gm,c} -
  \text{E}\left(\hat{d}_{gm,c}\right)\right)^4 =
\frac{3\pi^4}{16}\frac{d^4}{k^2} + O\left(\frac{1}{k^3}\right).
\end{align}

Therefore, we have completed the proof of Lemma \ref{lem_d_gm}.

\section{Proof of Lemma \ref{lem_d_gm_tail}}
\label{app_proof_lem_d_gm_tail} 

This section proves the tail bounds for $\hat{d}_{gm,c}$. 
Note that $\hat{d}_{gm,c}$ does not have a moment generating function
because  $\text{E}\left(\hat{d}_{gm,c}\right)^t=\infty$ if
$t\geq k$. However, we can still use the Markov moment bound.\footnote{In
fact, even when the moment generating function does exist, for any positive
random variable, the Markov moment bound is always sharper than the
Chernoff bound, although the Chernoff bound will be in an exponential
form. See \cite{Article:Philips_95,Article:Lugosi_04}.}  

For any $\epsilon \geq0$ and $0\leq t<k$, the Markov inequality says 
\begin{align}
\mathbf{Pr}\left(\hat{d}_{gm,c} \geq (1+\epsilon)d \right) \leq \frac{\text{E}\left(\hat{d}_{gm,c}\right)^t}{(1+\epsilon)^td^t}
= 
\frac{\cos^{kt}\left(\frac{\pi}{2k}\right)}{\cos^k\left(\frac{\pi
      t}{2k}\right)(1+\epsilon)^{t}},
\end{align}
\noindent which can be minimized by choosing the optimum $t = t_1^*$, 
where 
\begin{align}
t_1^* = \frac{2k}{\pi}\tan^{-1}\left(\left(\log(1+\epsilon) -
    k\log\cos\left(\frac{\pi}{2k}\right)\right)\frac{2}{\pi}\right). 
\end{align}

We need to make sure that $0\leq t_1^*<k$. $t_1^*\geq0$ because $\log\cos(.)\leq
0$; and $t_1^*<k$ because $\tan^{-1}(.) \leq \frac{\pi}{2}$, with
equality holding only when $k\rightarrow \infty$. 

For $0\leq \epsilon \leq1$, we can prove an exponential bound for 
$\mathbf{Pr}\left(\hat{d}_{gm,c} \geq (1+\epsilon)d \right)$. 
First of all, note that we do not
have to choose the optimum $t = t_1^*$. By the Taylor expansion, for
small $\epsilon$, $t_1^*$ can be well approximated by 
\begin{align}
t_1^* \approx \frac{4k\epsilon}{\pi^2} + \frac{1}{2} \approx
\frac{4k\epsilon}{\pi^2} = t_1^{**}.
\end{align}

Therefore, taking $t=t_1^{**} = \frac{4k\epsilon}{\pi^2}$, the tail bound becomes 
\begin{align}\notag
\mathbf{Pr}\left(\hat{d}_{gm,c} \geq (1+\epsilon)d \right) &\leq  \frac{\cos^{kt_1^{**}}\left(\frac{\pi}{2k}\right)}{\cos^k\left(\frac{\pi
      t_1^{**}}{2k}\right)(1+\epsilon)^{t_1^{**}}} \\\notag
&=
\left(\frac{\cos^{t_1^{**}}\left(\frac{\pi}{2k}\right)}{\cos\left(\frac{2\epsilon}{\pi}\right)(1+\epsilon)^{4\epsilon/\pi^2}}
\right)^k \\\notag
&\leq \left(\frac{1}{\cos\left(\frac{2\epsilon}{\pi}\right)(1+\epsilon)^{4\epsilon/\pi^2}}
\right)^k \\\notag
&=\exp\left(-k\left(\log\left(\cos\left(\frac{2\epsilon}{\pi}\right)\right) +
\frac{4\epsilon}{\pi^2}\log(1+\epsilon)\right)\right)
\\ 
&\leq \exp\left(-k\frac{\epsilon^2}{8(1+\epsilon)}\right),
\hspace{0.1in} 0\leq \epsilon\leq1\label{eqn_proof_right}
\end{align}

The last step in (\ref{eqn_proof_right}) needs some
explanations. First, by the Taylor expansion, 
\begin{align}\notag
&\log\left(\cos\left(\frac{2\epsilon}{\pi}\right)\right) +
\frac{4\epsilon}{\pi^2}\log(1+\epsilon) \\\notag
=& \left(-\frac{2\epsilon^2}{\pi^2} -
  \frac{4}{3}\frac{\epsilon^4}{\pi^4} +... \right)+
\frac{4\epsilon}{\pi^2}\left(\epsilon -
  \frac{1}{2}\epsilon^2+...\right)\\
=& \frac{2\epsilon^2}{\pi^2}\left(1-\epsilon+...\right)
\end{align}

Therefore, we can seek the smallest constant $\gamma_1$ so that
\begin{align}
\log\left(\cos\left(\frac{2\epsilon}{\pi}\right)\right) +
\frac{4\epsilon}{\pi^2}\log(1+\epsilon) 
\geq \frac{\epsilon^2}{\gamma_1(1+\epsilon)} =
\frac{\epsilon^2}{\gamma_1}(1-\epsilon +...)
\end{align}

\begin{figure}[h]
\begin{center}\mbox{
\subfigure[]{\includegraphics[width = 2.5in]{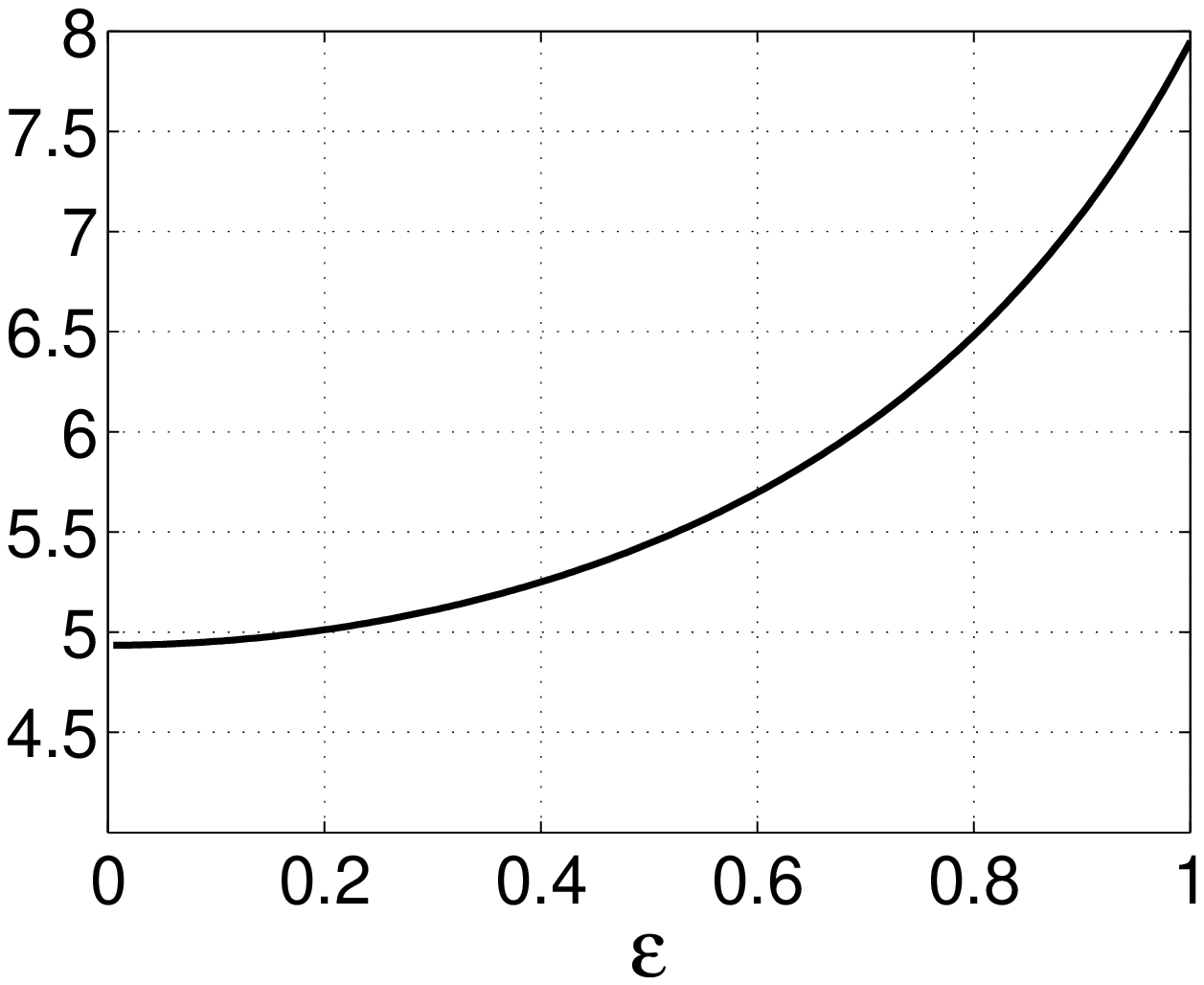}}\hspace{0.5in}
\subfigure[]{\includegraphics[width = 2.5in]{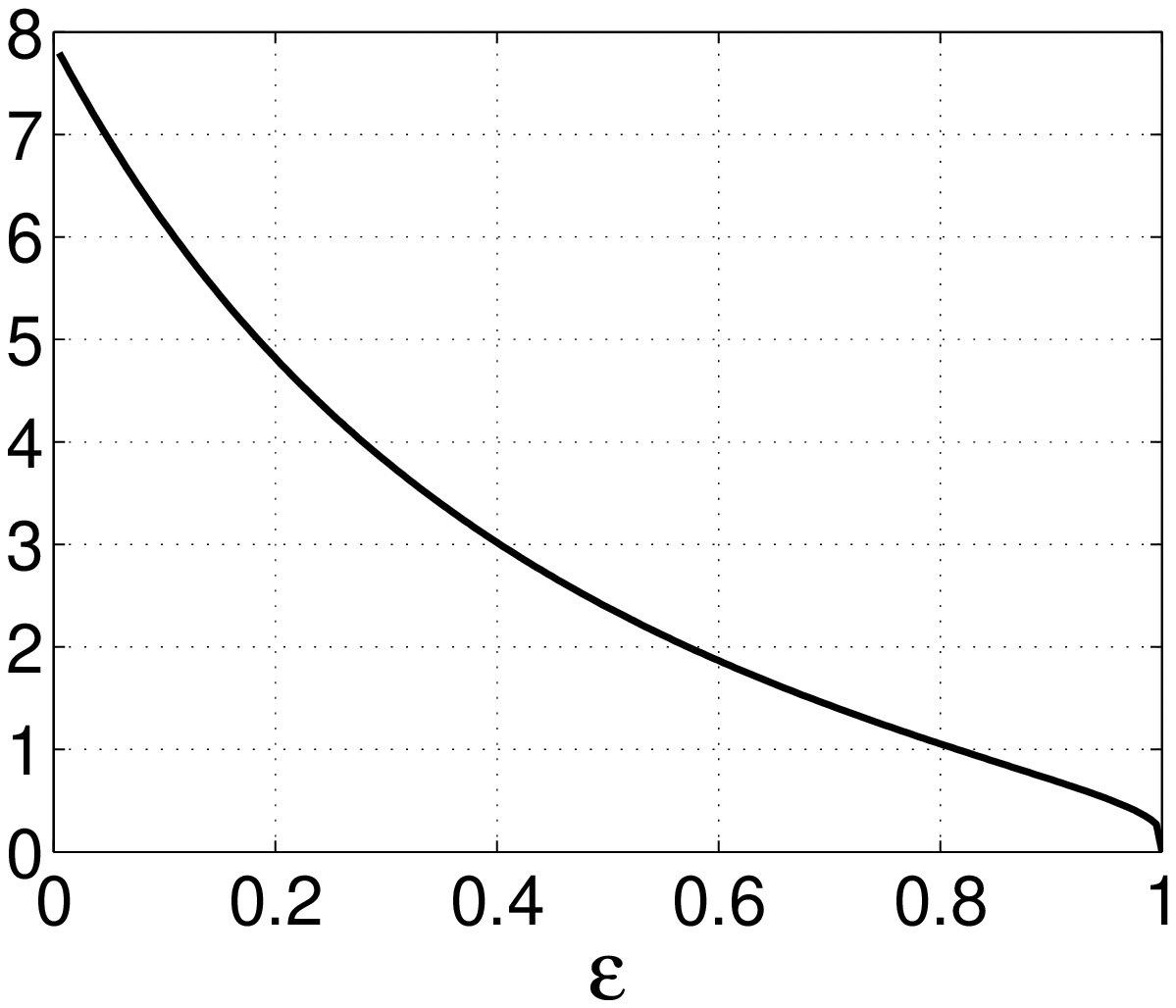}}}
\end{center}\vspace{-0.4in}
\caption{ (a):
  $\frac{\epsilon^2/(1+\epsilon)}{\log\left(\cos\left(\frac{2\epsilon}{\pi}\right)\right) +
\frac{4\epsilon}{\pi^2}\log(1+\epsilon)}$ as a function of
$\epsilon$. (b): $\frac{\epsilon^2/(1+\epsilon)}{\log\left(\cos\frac{2\epsilon}{\pi}\right) -
    \frac{4\epsilon}{\pi^2}\log(1-\epsilon) +
    \frac{8}{3}\log\left(\cos
      \frac{\epsilon}{3\pi}\right) }$ as a function of $\epsilon$. Graphically, we know that it suffices to use a constant 8
in (\ref{eqn_proof_right}) and (\ref{eqn_proof_left}). The optimal
constant will be different for different $\epsilon$. For example, if
$\epsilon = 0.2$, we could replace the constant 8 by a constant 5.  }\label{fig_gm_constant}\vspace{-0.2in}
\end{figure}

It is easy to see that as $\epsilon \rightarrow 0$,
$\gamma_1\rightarrow \frac{\pi^2}{2}$. Figure \ref{fig_gm_constant}(a)
illustrates that it suffices to let $\gamma_1 = 8$, which can be
numerically verified. This is why the last step in
(\ref{eqn_proof_right}) holds. Of course, we can get a better constant
if (e.g.,) $\epsilon =0.5$.

Now we need to  show the other tail bound $\mathbf{Pr}\left(\hat{d}_{gm,c}
  \leq  (1-\epsilon)d \right)$: 
\begin{align}\notag
&\mathbf{Pr}\left(\hat{d}_{gm,c} \leq  (1-\epsilon)d \right)  
=\mathbf{Pr}\left(\cos\left(\frac{\pi}{2k}\right)^k
  \prod_{j=1}^k|x_j|^{1/k} \leq  (1-\epsilon)d \right) \\\notag
=&\mathbf{Pr}\left(
  \sum_{j=1}^k\log\left(|x_j|^{1/k}\right)\leq
  \log\left(\frac{(1-\epsilon)d}{\cos^k\left(\frac{\pi}{2k}\right)}\right)\right)\\\notag
=&\mathbf{Pr}\left( \exp\left(
  \sum_{j=1}^k\log\left(|x_j|^{-t/k}\right)\right)\geq 
  \exp\left(-t\log\left(\frac{(1-\epsilon)d}{\cos^k\left(\frac{\pi}{2k}\right)}\right)\right)\right), \hspace{0.2in} 0\leq t<k \\
\leq & \left(\frac{(1-\epsilon)}{\cos^k\left(\frac{\pi}{2k}\right)}\right)^t
\frac{1}{\cos^k\left(\frac{\pi t}{2k}\right)}, \hspace{0.2in}
\text{(Chernoff bound)}
\end{align}
\noindent which is minimized at $t = t_2^*$ 
\begin{align}
t_2^* = \frac{2k}{\pi}\tan^{-1}\left(\left(-\log(1-\epsilon) +
    k\log\cos\left(\frac{\pi}{2k}\right)\right)\frac{2}{\pi}\right),
\end{align}
\noindent provided $k\geq \frac{\pi^2}{8\epsilon}$, otherwise $t_2^*$
may be less than 0. 

Again, $t_2^*$ can be replaced by its approximation 
\begin{align}
t_2^* \approx t_2^{**} = \frac{4k\epsilon}{\pi^2},  
\end{align}
\noindent provided $k\geq\frac{\pi^2}{4\epsilon}$, otherwise the
probability upper bound may exceed one.  Therefore, 

\begin{align}\notag
\mathbf{Pr}\left(\hat{d}_{gm,c} \leq  (1-\epsilon)d \right)  
\leq& \left(\frac{(1-\epsilon)}{\cos^k\left(\frac{\pi}{2k}\right)}\right)^{t_2^{**}}
\frac{1}{\cos^k\left(\frac{\pi t_2^{**}}{2k}\right)}\\\notag
=&\exp\left(-k\left(\log\left(\cos\frac{2\epsilon}{\pi}\right) -
    \frac{4\epsilon}{\pi^2}\log(1-\epsilon) +  \frac{4k\epsilon}{\pi^2}\log\left(\cos\frac{\pi}{2k}\right) \right)\right).
\end{align}
\noindent We can bound
$\frac{4k\epsilon}{\pi^2}\log\left(\cos\frac{\pi}{2k}\right)$ by restricting $k$.  

In order to attain $\mathbf{Pr}\left(\hat{d}_{gm,c} \leq  (1-\epsilon)d
\right)  \leq
\exp\left(-k\left(\frac{\epsilon^2}{8(1+\epsilon)}\right)\right)$, we
have to restrict $k$ to be larger than a certain value. For no
particular reason, we like to express the restriction as $k \geq
\frac{\pi^2}{\gamma_2\epsilon}$, for some constant $\gamma_2$. We
find $k \geq
\frac{\pi^2}{1.5\epsilon}$ suffices, although readers can verify that a
slightly better (smaller) restriction would be $k \geq
\frac{1}{4/\pi^2-1/4}\frac{1}{\epsilon} = \frac{\pi^2}{1.5326\epsilon} $. 

If $k \geq
\frac{\pi^2}{1.5\epsilon}$, then
$\frac{4k\epsilon}{\pi^2}\log\left(\cos\frac{\pi}{2k}\right) \geq
\frac{8}{3}\log\left(\cos \frac{\epsilon}{3\pi}\right)$. Therefore, 
\begin{align}\notag
\mathbf{Pr}\left(\hat{d}_{gm,c} \leq  (1-\epsilon)d \right) \leq  &\exp\left(-k\left(\log\left(\cos\frac{2\epsilon}{\pi}\right) -
    \frac{4\epsilon}{\pi^2}\log(1-\epsilon) +
    \frac{8}{3}\log\left(\cos
      \frac{\epsilon}{3\pi}\right)\right)\right)\\
\leq &\exp\left(-k\frac{\epsilon^2}{8(1+\epsilon)}\right), \hspace{0.2in}
k\geq \frac{\pi^2}{1.5\epsilon}\label{eqn_proof_left}
\end{align}

This completes the proof of Lemma \ref{lem_d_gm_tail}.

\section{Proof of Lemma \ref{lem_mle_asymp}} \label{app_proof_lem_asymp}
Assume $x \sim C(0,d)$. The $\log$
likelihood ($l(x;d)$) and first three derivatives
are  
\begin{align}
&l(x;d) = \log(d) - \log(\pi) - \log(x^2+d^2),\\
&l^\prime(d) = \frac{1}{d} - \frac{2d}{x^2+d^2}\\
&l^{\prime\prime}(d) = -\frac{1}{d^2} -
\frac{2x^2-2d^2}{(x^2+d^2)^2}\\
&l^{\prime\prime\prime}(d) = \frac{2}{d^3} +
\frac{4d}{(x^2+d^2)^2} + \frac{8d(x^2-d^2)}{(x^2+d^2)^3}
%\\
%&l^{\prime\prime\prime\prime}(d) = -\frac{6}{d^3} +
%\frac{4x^2-12d^2}{(x^2+d^2)^3} + \frac{8x^4-64x^2d^2+24d^4)}{(x^2+d^2)^4}.
\end{align}

The MLE  $\hat{d}_{MLE}$ is
asymptotically normal with mean $d$ and variance
$\frac{1}{k\text{I}(d)}$, where $\text{I}(d)$, the expected Fisher
Information, is 
\begin{align}
\text{I} = \text{I}(d) = \text{E}\left(-l^{\prime\prime}(d)\right)  =
\frac{1}{d^2} +
2\text{E}\left(\frac{x^2-d^2}{(x^2+d^2)^2}\right) = \frac{1}{2d^2},
\end{align}
\noindent because
\begin{align}\notag
\text{E}\left(\frac{x^2-d^2}{(x^2+d^2)^2}\right) &= \frac{d}{\pi}
\int_{-\infty}^\infty \frac{x^2-d^2}{(x^2+d^2)^3}dx \\ \notag
&=\frac{d}{\pi} \int_{-\pi/2}^{\pi/2} \frac{d^2(\tan^2(t) -
  1)}{d^6/\cos^6(t)} \frac{d}{\cos^2(t)}dt \\\notag
&=\frac{1}{d^2\pi}\int_{-\pi/2}^{\pi/2}\cos^2(t) - 2\cos^4(t) dt  \\
&= \frac{1}{d^2\pi}\left(\frac{\pi}{2}-2\frac{3}{8}\pi\right) = -\frac{1}{4d^2}
\end{align}
Therefore, we obtain
\begin{align}
\text{Var}\left(\hat{d}_{MLE}\right) = \frac{2d^2}{k} + O\left(\frac{1}{k^2}\right).
\end{align}

General formulas for the bias and higher moments of the MLE are
available in \citep{Article:Bartlett_53,Article:Shenton_63}.  We need to evaluate
the expressions in \cite[16a-16d]{Article:Shenton_63}, involving
tedious algebra: 
\begin{align}
&\text{E}\left(\hat{d}_{MLE}\right) = d - \frac{[12]}{2k\text{I}^2} +
O\left(\frac{1}{k^2}\right) \\
&\text{Var}\left(\hat{d}_{MLE}\right) = \frac{1}{k\text{I}} +
\frac{1}{k^2}\left(-\frac{1}{\text{I}}+\frac{[1^4]-[1^22]-[13]}{\text{I}^3}
+\frac{3.5[12]^2-[1^3]^2}{\text{I}^4}\right) + 
O\left(\frac{1}{k^3}\right) \\
&\text{E}\left(\hat{d}_{MLE}-\text{E}\left(\hat{d}_{MLE}\right)\right)^3 = 
\frac{[1^3]-3[12]}{k^2\text{I}^2}+O\left(\frac{1}{k^3}\right) \\\notag
&\text{E}\left(\hat{d}_{MLE}-\text{E}\left(\hat{d}_{MLE}\right)\right)^4= 
\frac{3}{k^2\text{I}^2} +
\frac{1}{k^3}\left(-\frac{9}{\text{I}^2}+
  \frac{7[1^4] - 6[1^22]-10[13]}{\text{I}^4}\right)\\
&\hspace{1.7in} +
\frac{1}{k^3}\left(\frac{-6[1^3]^2-12[1^3][12]+45[12]^2}{\text{I}^5}\right)+O\left(\frac{1}{k^4}\right),
\end{align}
\noindent where, after re-formatting,
\begin{align}\notag
&[12] = \text{E}(l^\prime)^3 +  \text{E}(l^\prime l^{\prime\prime}),
\hspace{0.3in} [1^4] = \text{E}(l^\prime)^4, \hspace{0.3in} [1^22] =
\text{E}(l^{\prime\prime}(l^\prime)^2) +  \text{E}(l^{\prime})^4, \\
&[13] = \text{E}(l^\prime)^4 +
3\text{E}(l^{\prime\prime}(l^\prime)^2)  + \text{E}(l^\prime
l^{\prime\prime\prime}), \hspace{0.3in} [1^3]=\text{E}(l^\prime)^3.
\end{align}

We will neglect most of the algebra. To help readers verifying the
results, the following formula we derive may be useful:
\begin{align}
\text{E}\left(\frac{1}{x^2+d^2}\right)^m =
\frac{1\times3\times5\times...\times(2m-1)}{2\times4\times6\times...\times(2m)}\frac{1}{d^{2m}},
\hspace{0.2in} m = 1, 2, 3, ...
\end{align}

Without giving the detail, we report 
\begin{align}\notag
&\text{E}\left(l^{\prime}\right)^3 = 0, \hspace{0.3in}
\text{E}\left(l^\prime l^{\prime\prime}\right) = -\frac{1}{2}\frac{1}{d^3}, \hspace{0.3in}
\text{E}\left(l^{\prime}\right)^4 =
\frac{3}{8}\frac{1}{d^4}, \\
&\text{E}(l^{\prime\prime}(l^\prime)^2) = -\frac{1}{8}\frac{1}{d^4},  \hspace{0.3in}
\text{E}\left(l^{\prime}l^{\prime\prime\prime}\right) =
\frac{3}{4}\frac{1}{d^4}.
\end{align}
Hence
\begin{align}
&[12] = -\frac{1}{2}\frac{1}{d^3}, \hspace{0.25in} [1^4] =
\frac{3}{8}\frac{1}{d^4}, \hspace{0.25in}[1^22] =
\frac{1}{4}\frac{1}{d^4}, \hspace{0.25in}[13] =
\frac{3}{4}\frac{1}{d^4}, \hspace{0.25in}[1^3] = 0.
\end{align}

Thus,  we obtain
\begin{align}
&\text{E}\left(\hat{d}_{MLE}\right) = d
+\frac{d}{k} + O\left(\frac{1}{k^2}\right)\\ 
&\text{Var}\left(\hat{d}_{MLE}\right) = \frac{2d^2}{k} + \frac{7d^2}{k^2} +
O\left(\frac{1}{k^3}\right) \\ 
&\text{E}\left(\hat{d}_{MLE}-\text{E}\left(\hat{d}_{MLE}\right)\right)^3 =
\frac{12d^3}{k^2} + O\left(\frac{1}{k^3}\right)\\
&\text{E}\left(\hat{d}_{MLE}-\text{E}\left(\hat{d}_{MLE}\right)\right)^4 =
\frac{12d^4}{k^2} + \frac{222d^4}{k^3} +  O\left(\frac{1}{k^4}\right).
\end{align}

Because $\hat{d}_{MLE}$ has $O\left(\frac{1}{k}\right)$ bias, we
recommend the bias-corrected estimator 
\begin{align}
\hat{d}_{MLE,c} = \hat{d}_{MLE}\left(1-\frac{1}{k}\right), 
\end{align}
whose first four moments are 
\begin{align}
&\text{E}\left(\hat{d}_{MLE,c}\right) = d + O\left(\frac{1}{k^2}\right)\\ 
&\text{Var}\left(\hat{d}_{MLE,c}\right) = \frac{2d^2}{k} + \frac{3d^2}{k^2} +
O\left(\frac{1}{k^3}\right) \\ 
&\text{E}\left(\hat{d}_{MLE,c}-\text{E}\left(\hat{d}_{MLE,c}\right)\right)^3 =
\frac{12d^3}{k^2} + O\left(\frac{1}{k^3}\right)\\
&\text{E}\left(\hat{d}_{MLE,c}-\text{E}\left(\hat{d}_{MLE,c}\right)\right)^4 =
\frac{12d^4}{k^2} + \frac{186d^4}{k^3} +  O\left(\frac{1}{k^4}\right),
\end{align}
\noindent by brute-force algebra. First, it is obvious that
\begin{align}
\text{E}\left(\hat{d}_{MLE} - d\right)^2 = \frac{2d^2}{k} + \frac{8d^2}{k^2}
+ O\left(\frac{1}{k^3}\right). 
\end{align}
Then 
\begin{align}\notag
\text{Var}\left(\hat{d}_{MLE,c}\right) &= \text{E}\left(\hat{d}_{MLE,c} -
  \text{E}(\hat{d}_{MLE,c})\right)^2\\\notag &=
\text{E}\left(\hat{d}_{MLE}\left(1-\frac{1}{k}\right) - d +
  O\left(\frac{1}{k^2}\right)\right)^2 \\\notag &= \text{E}\left(\left(\hat{d}_{MLE}-d\right)\left(1-\frac{1}{k}\right) -\frac{d}{k} +
  O\left(\frac{1}{k^2}\right)\right)^2  \\\notag 
&=\text{E}\left(\hat{d}_{MLE}-d\right)^2\left(1-\frac{2}{k}\right) +
\frac{d^2}{k^2} - 2\frac{d}{k}\left(1-\frac{1}{k}\right) +
O\left(\frac{1}{k^3}\right) \\
&= \frac{2d^2}{k} + \frac{3d^2}{k^2} + O\left(\frac{1}{k^3}\right). 
\end{align}

We can evaluate the higher central moments of $\hat{d}_{MLE,c}$ similarly,
but we skip the algebra.

Therefore, we have completed the proof for Lemma \ref{lem_mle_asymp}.

\end{document}